\documentclass[
reprint,
onecolumn,
nofootinbib,
amsmath,amssymb,
aps,
pra,
floatfix,
]{revtex4-2}

\usepackage{amsfonts}
\usepackage{mathtools}
\usepackage{bm}
\usepackage{array}[=2016-10-06]
\usepackage{dcolumn}
\usepackage{graphicx}
\usepackage{placeins}
\usepackage[usenames,dvipsnames]{xcolor}

\usepackage{hyperref}
\definecolor{citeblue}{RGB}{11, 61, 145}
\hypersetup{
	unicode=false,
	pdftitle={Tensor-network simulation of quantum phase estimation in the von Neumann measurement formulation with DMRG-prepared initial states},
	pdfauthor={Younes Javanmard},
	colorlinks=true,
	linkcolor=purple,
	citecolor=citeblue,
	urlcolor=teal
}

\graphicspath{ {./figures/} }

\providecommand{\ket}[1]{|#1\rangle}
\providecommand{\bra}[1]{\langle #1|}
\newcommand{\ra}{\rangle}
\newcommand{\la}{\langle}
\newcommand{\chidmrg}{\chi_{\mathrm{DMRG}}}
\newcommand{\chitdvp}{\chi_{\mathrm{TDVP}}}
\DeclareMathOperator{\Tr}{Tr}

\providecommand{\selectlanguage}[1]{}
\renewcommand{\selectlanguage}[1]{}

\begin{document}

\title{Tensor-network simulation of quantum phase estimation in the von Neumann measurement formulation with DMRG-prepared initial states}

\author{Younes Javanmard}
\email[]{javanmard.younes@gmail.com}
\affiliation{Institut f\"ur Theoretische Physik, Leibniz Universit\"at Hannover, Appelstra{\ss}e 2, 30167 Hannover, Germany}

\begin{abstract}
The von Neumann measurement formulation of quantum phase estimation [Phys.\ Rev.\ A \textbf{66}, 032314 (2002); Quantum \textbf{5}, 465 (2021)] estimates energies by coupling the system to a discretized pointer register through $\hat H\otimes\hat p$ and reading out the pointer after an inverse quantum Fourier transform. Building on these works, we develop a tensor-network framework for simulating and benchmarking this protocol when the system is initialized in a matrix product state (MPS) pre-trained with the density matrix renormalization group (DMRG). The system and the pointer are represented as a composite MPS, the coupling $\hat H\otimes\hat p$ as a matrix product operator, and the coupled dynamics is evolved with the time-dependent variational principle. This gives direct access to the time-resolved pointer distribution, the recovered energy, and the entanglement generated between system and pointer. We apply the framework to quantum many-body spin systems and to electronic-structure problems. For the systems studied, the pointer readout yields accurate ground-state energy estimates from DMRG-prepared approximate states, even when their variational energies are not fully converged. These states generally contain contributions from both ground and excited eigenstates, with their spectral weights determining the corresponding contributions to the pointer distribution. More accurate initial states also strongly suppress the entanglement generated between system and pointer. We also provide logical and compiled resource estimates for a Suzuki--Trotter realization of the protocol.
\end{abstract}

\maketitle

\section{Introduction}
\label{sec:intro}

One key application of emerging quantum computers is the quantum simulation of complex many-body systems, with prospective impact on the development of new materials and industrial processes, for example in battery research and drug discovery. The efficient simulation of quantum many-body systems is computationally demanding because of the tensor-product structure of the underlying Hilbert space. Tensor networks (TN)~\cite{RevModPhys.93.045003, Astrakhantsev23, bridgemanHandwavingInterpretiveDance2017, biamonte2017tensor} provide a parsimonious description of the collective and emergent behavior of such systems. The primary example is the density matrix renormalization group (DMRG)~\cite{whiteDensityMatrixFormulation1992, schollwockDensitymatrixRenormalizationGroup2005, schollwockDensitymatrixRenormalizationGroup2011}, which efficiently finds ground states of one-dimensional Hamiltonians by minimizing $\la\psi|\hat H|\psi\ra$ over matrix product states (MPS); its precision is controlled by the maximal bond dimension, which also sets its computational cost. Tensor networks can, however, encounter limitations for highly entangled states, non-local interactions, higher-dimensional systems, long-time real-time dynamics, and when very high numerical precision is required.

Quantum algorithms such as quantum phase estimation (QPE)~\cite{Kitaev1995QuantumMA, AbramsPRL1999, Nielsen_Chuang_2010} and quantum singular value transformation can in principle provide highly accurate ground-state energies~\cite{Nielsen_Chuang_2010, Portugal_quantum_2013}, but a demonstration of quantum advantage with these methods requires fault-tolerant quantum processing units (QPUs). Hybrid classical--quantum workflows, which combine high-performance classical computing with QPUs, may therefore become increasingly useful for quantum simulation~\cite{preskillQuantumComputingNISQ2018, peruzzoVariationalEigenvalueSolver2014, farhiQuantumApproximateOptimization2014}. The logical resource estimates presented in this work should accordingly be read as forward-looking compilation estimates rather than as evidence of immediate executability on present-day noisy hardware.

A central difficulty of quantum algorithms for low-energy physics is the choice of the initial state: the probability that phase estimation projects onto the ground state equals the squared overlap of the input with that ground state, so state preparation is often the bottleneck. A natural strategy, analogous to the use of pre-trained models in machine learning, is to start from a classically pre-optimized (``pre-trained'') solution~\cite{dborinMatrixProductState2022, huangTensorNetworkAssisted2022, khanPreoptimizingVariationalQuantum2023, shinAnalyzingQuantumMachine2023, rieserTensorNetworksQuantum2023, fanQuantumCircuitMatrix2023, termanova2024tensor, robledomoreno2024chemistry}. Tensor-network methods are well suited for this role: DMRG provides an MPS approximation of the ground state, and an MPS can be loaded into a quantum register by a sequential circuit~\cite{schoenSequentialGenerationMatrixProduct2007, Ran20, Lin21, rudolph2022decomposition, Dborin_2022, Astrakhantsev23, YJ_mps_ansatz2024}.

In this paper we study such a pre-trained MPS as the input of one particular form of phase estimation, the \emph{von Neumann measurement formulation}. This formulation is not new, and we first describe the two works on which our study is built.

\paragraph*{Childs \emph{et al.} (2002).} In ``Quantum search by measurement'', Childs \emph{et al.}~\cite{Qsearchmeasurement2002} showed how von Neumann's model of an energy measurement~\cite{vonNeumann1955} can be implemented on a quantum computer. The continuous pointer is replaced by $r$ qubits with discretized momentum $\hat p=\sum_{j=1}^{r}2^{-j}(1-\hat Z_j)/2$; the pointer is coupled to the system through $e^{-it\hat H\otimes\hat p}$ and read out after an inverse quantum Fourier transform (QFT). They noted that this circuit is a form of phase estimation, and used sequences of such measurements of slowly varying Hamiltonians to keep the system in the instantaneous ground state---a measurement-based analogue of adiabatic evolution---recovering the quadratic speed-up of Grover's search. The circuit of Fig.~\ref{fig:circuit}, the pointer operator, and the readout rule used in this paper are all taken from Ref.~\cite{Qsearchmeasurement2002}.

\paragraph*{Novo \emph{et al.} (2021).} In ``Quantum advantage from energy measurements of many-body quantum systems'', Novo \emph{et al.}~\cite{novo_quantum_2021} applied the same finite-precision energy measurement to local many-body Hamiltonians. They provided an error analysis of the finite pointer and finite evolution time, and showed that sampling from the resulting energy-outcome distribution is classically hard under plausible complexity-theoretic conjectures, even for simple input states, proposing energy measurements as a route to demonstrating quantum advantage. Our error discussion in Sec.~\ref{sec:precision} and Appendix~\ref{app:vN_error_analysis} follows their analysis. More recently, Skelton~\cite{Skelton2026vonNeumannQPE} combined von Neumann's measurement procedure with Hamiltonian simulation of block-encoded Hamiltonians and derived Clifford+T complexity bounds for quantum phase and energy estimation, starting from a linear combination of Pauli strings, arguing that the approach remains practical for fault-tolerant algorithms.

\paragraph*{This work.} We take the protocol exactly as defined in Refs.~\cite{Qsearchmeasurement2002, novo_quantum_2021} and address a question those works do not consider: \emph{what does the pointer readout look like, and how accurate is the recovered energy, when the input is a DMRG-pretrained MPS of small bond dimension for a realistic spin or molecular Hamiltonian?} Answering this requires simulating the coupled system--pointer dynamics, which we do with tensor networks: the system and the finite pointer register are represented as one composite MPS, the coupling $\hat H\otimes\hat p$ as an MPO, and the composite state is evolved with the time-dependent variational principle (TDVP)~\cite{haegeman_time-dependent_2011, haegeman_unifying_2016}. This makes it possible to study the accuracy, spectral resolution, and initial-state dependence of the energy readout before moving to explicit compiled-circuit simulations. Table~\ref{tab:prior_work} summarizes how the three works relate.

\begin{table}[t]
\caption{Comparison of the two works that introduced and analyzed the von Neumann measurement formulation of phase estimation with the present paper. The quantum protocol is the same in all three. The present work combines DMRG-prepared inputs with tensor-network simulation and numerical benchmarks of this established protocol.}
\label{tab:prior_work}
\begin{ruledtabular}
\begin{tabular}{>{\raggedright\arraybackslash\bfseries}p{0.13\textwidth} >{\raggedright\arraybackslash}p{0.22\textwidth} >{\raggedright\arraybackslash}p{0.22\textwidth} >{\raggedright\arraybackslash}p{0.33\textwidth}}
 & Childs \emph{et al.} (2002)~\cite{Qsearchmeasurement2002} & Novo \emph{et al.} (2021)~\cite{novo_quantum_2021} & This work \\
\hline
Goal & Solve unstructured search (and adiabatic-type problems) by a sequence of measurements & Establish conditional hardness of classical energy sampling & Benchmark how accurately the protocol recovers ground-state energies from DMRG-pretrained inputs \\[2pt]
Quantum protocol & Introduced: discretized von Neumann pointer, $e^{-it\hat H\otimes\hat p}$, inverse-QFT readout & Same protocol, applied to local many-body Hamiltonians & Same protocol, adopted unchanged from Refs.~\cite{Qsearchmeasurement2002, novo_quantum_2021} \\[2pt]
Input state & States along an interpolating Hamiltonian path (search problem) & Simple (e.g.\ product) states & DMRG-optimized MPS of small bond dimension; sequential loading circuit described \\[2pt]
Hamiltonians & Search/adiabatic Hamiltonians & Local lattice Hamiltonians & Frustrated quantum spin models and molecular electronic-structure Hamiltonians \\[2pt]
Method of analysis & Analytic (run-time and success-probability bounds) & Analytic (error bounds, complexity-theoretic hardness) & Numerical: general tensor-network pipeline (system+pointer MPS, MPO for $\hat H\otimes\hat p$, TDVP) for any Hamiltonian with an MPO representation; ideal protocol \\[2pt]
Main output & Measurement-based algorithm achieving the Grover speed-up & Hardness of classically sampling energy distributions & Time-resolved pointer distributions, fitted energies, system--pointer entanglement vs.\ initial-state quality \\[2pt]
Circuit resources & Not the focus & Not the focus & Logical CNOT counts for a Trotterized realization and counts after compilation with available compilers (Qiskit, Rustiq, TKET); pointer overhead and connectivity considerations \\
\end{tabular}
\end{ruledtabular}
\end{table}

\paragraph*{Main results.} In this work, we provide:
\begin{enumerate}
\item \emph{A tensor-network framework for the coupled system--pointer dynamics} (Sec.~\ref{sec:tn}). The system and pointer form one MPS, $\hat H\otimes\hat p$ is constructed as an MPO with the finite-state-machine (matrix-product-diagram) method~\cite{crosswhite2008, Keller2015MPO, Ren2020, javanmard2026coefficientdecoupledmatrixproductoperators}, and the evolution uses two-site TDVP. The pointer distribution is computed exactly from the evolved MPS, so no sampling noise enters the benchmark; the effect of a finite number of measurements is quantified separately by perfect sampling of the MPS (Appendix~\ref{app:sampling}).
\item \emph{Benchmarks with DMRG-prepared initial states} (Sec.~\ref{sec:benchmarks}). For a frustrated 12-site Heisenberg model and three molecular Hamiltonians of 13--16 qubits, we report the full time-resolved pointer distribution, the energy recovered from it, and its error relative to exact or FCI references, and we show that the system--pointer entanglement is strongly suppressed when the DMRG input is more accurate.
\item \emph{Auxiliary logical and compiled resource estimates} (Sec.~\ref{sec:resources}) for a first-order Suzuki--Trotter realization of the benchmark instances, with the controlled rotations required by the pointer coupling accounted for explicitly, and gate counts after compilation with current circuit compilers.
\end{enumerate}
The benchmarks concern the ideal (noise-free, Trotter-error-free) protocol simulated with TDVP; the Suzuki--Trotter construction is used only as an auxiliary implementation model for gate counting. Our approach can also be regarded as a way to simulate phase-estimation circuits with a TDVP-based tensor-network method, and is complementary to recent work on TDVP-based simulation of explicit compiled quantum circuits~\cite{sander2025quantumcircuitsimulationlocal}. Its relation to hybrid tensor-network quantum simulation~\cite{Yuan2021PRLHybridTN, Yuan2020arxivHybridTN} is discussed in Sec.~\ref{sec:relation_hybrid_tn}.

\paragraph*{Outline.} Section~\ref{sec:background} reviews the von Neumann measurement protocol of Refs.~\cite{Qsearchmeasurement2002, novo_quantum_2021} and the DMRG/MPS state-preparation input; it contains no new results. Section~\ref{sec:tn} presents the tensor-network simulation framework. Section~\ref{sec:benchmarks} presents the benchmarks for spin systems and electronic-structure problems. Section~\ref{sec:resources} gives the auxiliary resource estimates, Sec.~\ref{sec:discussion} discusses the scope and limitations of the results, and Sec.~\ref{sec:conclusion} concludes. The appendices contain the Jordan--Wigner mapping, the graph-based MPO construction for molecular Hamiltonians, molecular geometries, an additional molecular benchmark ($\mathrm{BH}_3$), details of the TDVP scheme, the idealized error analysis of the pointer readout, the finite-shot readout by perfect sampling of the MPS, a discussion of product-formula error, and a remark on excited states.

\section{Background: the von Neumann measurement formulation of QPE}
\label{sec:background}
This section reviews material from Refs.~\cite{Qsearchmeasurement2002, novo_quantum_2021, vonNeumann1955, Mello2014, Temme2011, VermershPRX2019} in the notation used in the rest of the paper. The protocol has three steps: preparation of the initial state (here by DMRG, Sec.~\ref{sec:mps_loading}), the coupled system--pointer evolution with readout (Secs.~\ref{sec:pointer}--\ref{sec:readout}), and classical post-processing of the pointer distribution.

\subsection{Hamiltonians}
In practice, the Hamiltonian of interest is written as a sum of $K$ Pauli strings,
\begin{equation}
\hat H=\sum_{i=1}^{K}c_i\,\hat h_i,
\label{eq:H_pauli}
\end{equation}
where each term is an $N$-qubit Pauli string
\begin{equation}
\hat h_i=\bigotimes_{j=1}^{N}\hat\sigma^{(i)}_j,\qquad \hat\sigma^{(i)}_j\in\{\hat I,\hat X,\hat Y,\hat Z\},
\end{equation}
and the superscript $(i)$ labels the Pauli operator on site $j$ in the $i$-th term. Throughout, $K$ denotes the number of Pauli terms, $\hbar=1$, and the eigenpairs of $\hat H$ are $\hat H\ket{\psi_j}=E_j\ket{\psi_j}$, with ground state $\ket{\Omega}\equiv\ket{\psi_0}$ and ground-state energy $E_0$. We are interested in the ground-state energy of such Hamiltonians.

\subsection{Discretized von Neumann measurement}
\label{sec:pointer}
The von Neumann prescription provides a general framework for measurements in quantum mechanics, applicable to all observables and systems~\cite{vonNeumann1955, Mello2014}. The system, in the input state $\ket\psi$, is coupled to an ancillary subsystem representing a continuous degree of freedom---a free particle on a line or a harmonic oscillator---initialized in the position ``eigenstate'' $\ket{x=0}$. This ancilla acts as a \emph{meter}: its position becomes correlated with the energy of the system, the particle moving further to the right the higher the energy. A measurement of the meter's position then collapses it onto the position eigenstate corresponding to the outcome~\cite{vonNeumann1955, Mello2014, Qsearchmeasurement2002, novo_quantum_2021, VermershPRX2019}.

\begin{figure}[t]
\centering
\includegraphics[width=0.62\textwidth]{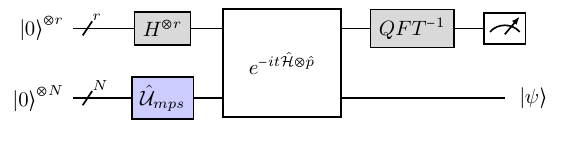}
\caption{Circuit of the von Neumann measurement formulation of phase estimation~\cite{Qsearchmeasurement2002, novo_quantum_2021}, with a pre-trained MPS as input. Top: $r$ pointer qubits, prepared in $\ket{x=0}$ [Eq.~\eqref{eq:pointer_init}] by Hadamard gates $H^{\otimes r}$. Bottom: $N$ system qubits, prepared in the DMRG-optimized MPS approximating the ground state by the loading circuit $\hat{\mathcal U}_{\mathrm{mps}}$ (Sec.~\ref{sec:mps_loading}). The two registers evolve jointly under $e^{-it\hat H\otimes\hat p}$ [Eq.~\eqref{eq:sys_pointer_te}]; after a time $t$ an inverse QFT is applied to the pointer, which is then measured in the computational basis. Repeating the circuit gives the distribution $P(x;t)$ of Eq.~\eqref{eq:Pxt}, which is post-processed to estimate the ground-state energy. In Secs.~\ref{sec:tn}--\ref{sec:benchmarks} this circuit is simulated with tensor networks, with $r=4$ in all benchmarks.}
\label{fig:circuit}
\end{figure}

The energy of a Hermitian operator can only be measured to finite precision $\epsilon$ with a finite quantum circuit, and the circuit depth grows with the required precision. Following Childs \emph{et al.}~\cite{Qsearchmeasurement2002}, the circuit is obtained by discretizing von Neumann's prescription; in one of its manifestations it is phase estimation~\cite{Mello2014, Temme2011}. The circuit is shown in Fig.~\ref{fig:circuit}.

The continuous pointer is discretized with $r$ qubits, replacing the continuous variable by a $2^r$-dimensional space. The computational basis states $\ket z$ of the pointer are the momentum eigenstates, labeled by the integers $z=0,\dots,2^r-1$ in binary. The discretized momentum operator is the sum of single-qubit terms $\hat p_j$,
\begin{equation}
\hat p=\sum_{j=1}^{r}2^{-j}\,\frac{1-\hat Z_j}{2}=\sum_{j=1}^{r}\hat p_j,
\label{eq:pointer}
\end{equation}
normalized such that
\begin{equation}
\hat p\ket{z}=\frac{z}{2^r}\ket{z}.
\end{equation}
For example, for $r=2$ there are $2^2=4$ states, $\ket{00}\equiv\ket0$, $\ket{10}\equiv\ket1$, $\ket{01}\equiv\ket2$, and $\ket{11}\equiv\ket3$.

The pointer is initialized in the discretized position state $\ket{x=0}$, a narrow wave packet centered at $x=0$,
\begin{equation}
\ket{x=0}=\frac{1}{2^{r/2}}\sum_{z=0}^{2^r-1}\ket{z},
\label{eq:pointer_init}
\end{equation}
which is prepared from $\ket0^{\otimes r}$ by a layer of Hadamard gates (equivalently, a QFT). Since the momentum generates translations, the displacement operator $\hat D_\alpha=e^{-i\alpha\hat p}$ shifts the pointer,
\begin{equation}
e^{-i\alpha\hat p}\ket{x=0}=\ket{x=\alpha}.
\end{equation}

The system is coupled to the pointer through $\hat H\otimes\hat p$, which for the Hamiltonian of Eq.~\eqref{eq:H_pauli} reads
\begin{equation}
\hat H\otimes\hat p=\sum_{i=1}^{K}\sum_{j=1}^{r}c_i\,\hat h_i\otimes\hat p_j .
\label{eq:coupling}
\end{equation}
If $\hat H$ contains at most $k$-body terms, $\hat H\otimes\hat p$ contains at most $(k+1)$-body terms, so its dynamics can be simulated with the same methods as that of $\hat H$. Since $\hat H\otimes\hat p$ is block diagonal in the eigenbasis of $\hat H$, the joint evolution for a time $t$ is
\begin{equation}
e^{-it\hat H\otimes\hat p}=\sum_{j}\ket{\psi_j}\!\bra{\psi_j}\otimes e^{-itE_j\hat p}.
\label{eq:sys_pointer_te}
\end{equation}
If the system is in the eigenstate $\ket{\psi_j}$, the pointer is displaced from $x=0$ to $x=tE_j$,
\begin{equation}
e^{-it\hat H\otimes\hat p}\ket{\psi_j}\ket{x=0}=\ket{\psi_j}\ket{x=tE_j},
\end{equation}
or, in the pointer's computational basis,
\begin{equation}
e^{-it\hat H\otimes\hat p}\ket{\psi_j}\ket{x=0}=\frac{1}{2^{r/2}}\sum_{z=0}^{2^r-1}e^{-iE_jzt/2^r}\ket{\psi_j}\ket{z}.
\label{eq:coupled_evolution}
\end{equation}

\subsection{Pointer-readout distribution}
\label{sec:readout}
Applying the inverse QFT to the pointer leaves the system and pointer in $\ket{\psi_j}\otimes\ket\phi$ with
\begin{equation}
\ket\phi=\sum_{x=0}^{2^r-1}\Big(\frac{1}{2^r}\sum_{z=0}^{2^r-1}e^{-\frac{2\pi i}{2^r}\left(x+\frac{E_jt}{2\pi}\right)z}\Big)\ket{x}
=\sum_{x=0}^{2^r-1}f(E_j,x)\ket{x},
\end{equation}
where
\begin{equation}
|f(E_j,x)|^2=\frac{1}{4^r}\,\frac{\sin^2\!\big(\pi(x+E_jt/2\pi)\big)}{\sin^2\!\big(\tfrac{\pi}{2^r}(x+E_jt/2\pi)\big)},
\label{eq:pointerDist}
\end{equation}
which is strongly peaked at $x^\star\equiv-E_jt/2\pi\pmod{2^r}$; for the negative ground-state energies considered here, the peak therefore moves to larger $x$ as $t$ increases. For a general input $\ket{\psi}=\sum_j a_j\ket{\psi_j}$, the outcome distribution is the mixture
\begin{equation}
P(x;t)=\sum_j|a_j|^2\,|f(E_j,x)|^2,\qquad |a_j|^2=|\la\psi_j|\psi\ra|^2 ,
\label{eq:Pxt}
\end{equation}
so the ground-state peak carries the weight $|\la\Omega|\psi\ra|^2$. This is why a high-overlap initial state is essential. At each evolution time one may therefore measure the pointer and estimate the energy from the distribution~\eqref{eq:pointerDist}; Ref.~\cite{novo_quantum_2021} provides a detailed error analysis of this procedure.

\paragraph*{Time-resolved view used in our figures.} In the benchmarks we display $P(x;t)$ for each of the $2^r$ outcomes $x$ as a function of $t$. For an input dominated by $\ket{\Omega}$, the peak moves through the outcomes as $t$ increases: it advances by one outcome every $\Delta t=2\pi/|E_0|$ and returns to the same outcome after the revival time
\begin{equation}
T_{\mathrm{rev}}=2^r\,\frac{2\pi}{|E_0|}.
\label{eq:revival}
\end{equation}
Each row $x$ of the figures therefore shows a train of peaks spaced by $T_{\mathrm{rev}}$, and consecutive rows are shifted by $\Delta t$. Fitting the time trace of row $x$ to Eq.~\eqref{eq:pointerDist}, with $E$ as the only free parameter (Sec.~\ref{sec:tn_readout}), gives an energy estimate $E_{\mathrm{es}}(x)$, so the $2^r$ rows provide $2^r$ estimates of the same energy; the figures report each $E_{\mathrm{es}}(x)$ together with their median $E_{\mathrm{estimated}}$.

\subsection{Idealized precision guarantees for the pointer readout}
\label{sec:precision}
We summarize the idealized precision analysis; the derivation, a gap-resolvability condition, and a worked chemical-accuracy example are given in Appendix~\ref{app:vN_error_analysis}, following Ref.~\cite{novo_quantum_2021}. The $r$-qubit readout produces a distribution $P(x)$ over $x\in\{0,\dots,2^r-1\}$ peaked near $x^\star\approx-Et/2\pi\pmod{2^r}$ when the system is in an eigenstate of energy $E$. Given a measured or fitted peak location $\hat x$, we use the estimator
\begin{equation}
\hat E=-\frac{2\pi\,\hat x}{t}.
\label{eq:E_hat}
\end{equation}

\paragraph*{Aliasing / energy-range condition.} Since the readout is defined modulo $2^r$, wrap-around is avoided if the relevant energy range fits into the phase window; a sufficient condition is
\begin{equation}
\frac{(E_{\max}-E_{\min})\,t}{2\pi}<2^r,
\label{eq:no_aliasing}
\end{equation}
where $[E_{\min},E_{\max}]$ is a known spectral interval of interest (or an upper bound on it).

\paragraph*{Single-shot success guarantee and accuracy.} For an integer acceptance window of half-width $w>1$ around the peak, $|\hat x-x^\star|\le w$, a standard phase-estimation tail bound gives
\begin{equation}
\Pr(\text{success})\ge1-\frac{1}{2(w-1)}\equiv1-\delta,
\label{eq:success_prob}
\end{equation}
so the single-shot failure probability is at most $\delta$. Conditioned on success, the estimator obeys
\begin{equation}
|E-\hat E|\le\frac{2\pi w}{t}.
\label{eq:energy_error}
\end{equation}

\paragraph*{Sampling (shot) noise.} Let $p_x=P(x)$ be the ideal probability of outcome $x$ and $\hat p_x$ its empirical estimate from $M$ circuit repetitions. Then
\begin{equation}
\mathrm{Std}(\hat p_x)=\sqrt{\frac{p_x(1-p_x)}{M}}\le\frac{1}{2\sqrt M}.
\end{equation}
These fluctuations induce an uncertainty in the inferred peak position $\hat x$ and hence in $\hat E=-2\pi\hat x/t$; for a fixed peak shape one expects~\cite{WassermanAllStats2004}
\begin{equation}
\mathrm{Std}(\hat E)\sim\mathcal O\!\left(\frac{1}{t\sqrt M}\right),
\end{equation}
with a prefactor that depends on the peak-finding or fitting procedure. The benchmarks in Sec.~\ref{sec:benchmarks} use the exact distribution $P(x;t)$ and therefore contain no shot noise.

\paragraph*{Hamiltonian simulation and hardware noise.} On a digital quantum computer the evolution would be implemented with a Suzuki--Trotter formula with $t=n\delta t$; reducing $\delta t$, or using higher-order formulas, decreases the simulation error at the cost of circuit depth. At finite $\delta t$, the product-formula circuit implements evolution under an effective Hamiltonian $\widetilde H_{\delta t}$ that differs from $\hat H$ by terms that vanish as $\delta t\to0$, so the energy inferred from the pointer may be shifted relative to the exact eigenvalue (Appendix~\ref{app:trotter_error_analysis}). The TDVP/MPS benchmarks of this work do not quantify this Trotter-induced shift. We also do not perform device-level noise simulations; qualitatively, two-qubit-gate infidelity, routing overhead for long-range controls, and readout errors reduce the visibility of the peaks in $P(x)$ and thereby increase the uncertainty of the fitted energy.

\subsection{Relation to textbook QPE}
\label{sec:vs_qpe}
As noted in Ref.~\cite{Qsearchmeasurement2002}, the von Neumann measurement of $\hat H=i\log U$ is equivalent to phase estimation of $U=e^{-i\hat Ht}$~\cite{Nielsen_Chuang_2010}. Indeed, since $\hat p_j$ acts on pointer qubit $j$ only, $e^{-it\hat H\otimes\hat p}=\prod_j e^{-it2^{-j}\hat H\otimes(1-\hat Z_j)/2}$ is a product of evolutions controlled by the individual pointer qubits. Typical textbook QPE implementations~\cite{Kitaev1995QuantumMA, AbramsPRL1999, Nielsen_Chuang_2010} apply a sequence of controlled powers $U^{2^k}$, which can give high precision with relatively few repetitions at the cost of a large coherent depth per run. In the von Neumann formulation used here, the same controlled evolution is written as a single evolution of the system and pointer under $\hat H\otimes\hat p$ per shot, and the energy is reconstructed from repeated sampling of the pointer distribution together with classical post-processing. This leads to a different trade-off between coherent circuit depth, number of repetitions, pointer-register size, and classical inference. We do not claim that this trade-off is superior to standard QPE or to modern fault-tolerant spectral-estimation methods. We use the von Neumann form because the coupled evolution can be simulated directly as a Hamiltonian time evolution with TDVP (Sec.~\ref{sec:tn}) and compiled directly with product formulas (Sec.~\ref{sec:resources}). Both formulations require a non-negligible overlap of the input with the target eigenstate [Eq.~\eqref{eq:Pxt}].

\subsection{State preparation using DMRG}
\label{sec:mps_loading}
The key classical input of the workflow is White's DMRG~\cite{whiteDensityMatrixFormulation1992, schollwockDensitymatrixRenormalizationGroup2005, schollwockDensitymatrixRenormalizationGroup2011}, a variational method that represents the ground state $\ket\Omega$ as an MPS~\cite{fannesFinitelyCorrelatedPure1994, verstraeteMatrixProductStates2006, hastingsAreaLawOnedimensional2007, schollwockDensitymatrixRenormalizationGroup2011},
\begin{equation}
\ket{\psi}=\sum_{\substack{\sigma_1\ldots\sigma_N\\ \alpha_1\ldots\alpha_{N-1}}}T^{\sigma_1}_{\alpha_0\alpha_1}\cdots T^{\sigma_i}_{\alpha_{i-1}\alpha_i}\cdots T^{\sigma_N}_{\alpha_{N-1}\alpha_N}\ket{\sigma_1\cdots\sigma_i\cdots\sigma_N},
\label{eq:MPS}
\end{equation}
where each $T^{\sigma_i}_{\alpha_{i-1}\alpha_i}$ is a rank-3 tensor, the bond indices $\alpha_i$ run from 1 to the bond dimension $\chi_i$, and $\chi_0=\chi_N=1$ (open boundary conditions). We denote the maximal bond dimension of the DMRG state by $\chidmrg$. DMRG sequentially optimizes the tensors, carrying out
\begin{equation}
\min_{T^{\sigma_i}_{\alpha_{i-1}\alpha_i}}\la\psi|\hat H|\psi\ra
\end{equation}
for each site $i$ and iterating until convergence. MPS provide faithful representations of ground states of gapped one-dimensional systems~\cite{verstraeteMatrixProductStates2006, hastingsAreaLawOnedimensional2007}, which underlies the general applicability of DMRG. In this work $\chidmrg$ is deliberately kept small, so that the input is only an approximation of $\ket\Omega$.

\begin{figure}[t]
\centering
\includegraphics[width=0.85\textwidth]{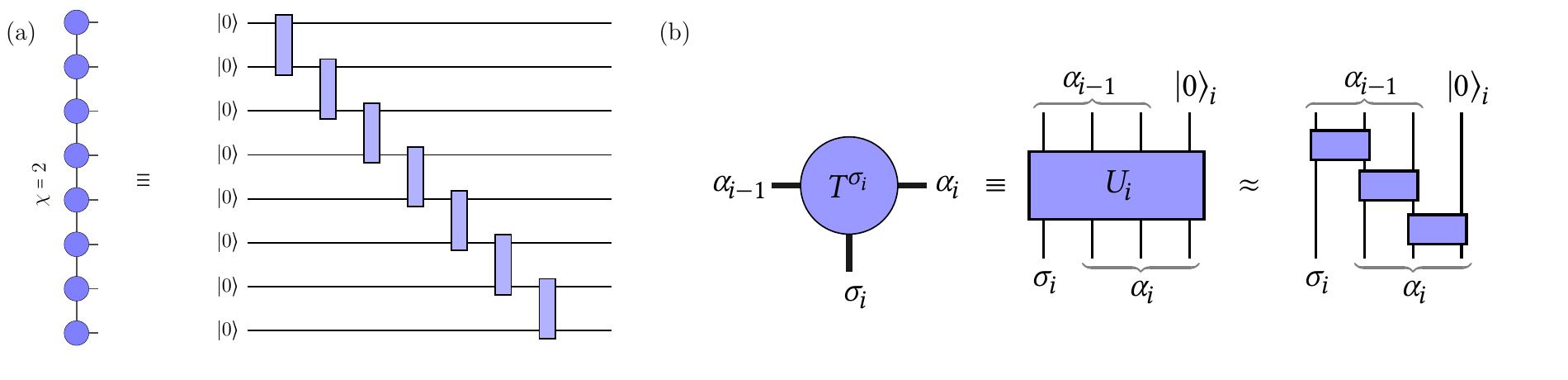}
\caption{Standard sequential circuit for loading an MPS into a quantum register~\cite{schoenSequentialGenerationMatrixProduct2007, Ran20}, drawn for bond dimension $\chi=2$. (a)~MPS in right-canonical form. (b)~Equivalent staircase circuit: each isometric tensor $T^{\sigma_i}$ becomes a local two-qubit unitary $U_i$ [Eq.~\eqref{eq:isometry}] applied sequentially along the chain. For larger bond dimensions the same sequential structure applies with $U_i$ acting on $1+\lceil\log_2\chi\rceil$ qubits; compiling these unitaries into a native gate set gives a depth that depends on the compilation strategy and increases with $\chi$. The $\chi=2$ case is included only to illustrate the loading principle; the bond dimensions used in the benchmarks are given in Table~\ref{tab:benchmarks}.}
\label{fig:mps_loading}
\end{figure}

To load an MPS into a quantum register~\cite{schoenSequentialGenerationMatrixProduct2007, Ran20, Lin21, rudolph2022decomposition, Dborin_2022, Astrakhantsev23, YJ_mps_ansatz2024}, it is first brought into right-canonical form~\cite{schollwockDensitymatrixRenormalizationGroup2011, javanmard_sharp_2018} using the gauge freedom of the MPS,
\begin{equation}
\sum_{\sigma_i,\alpha_i}T^{\sigma_i}_{\alpha_{i-1}\alpha_i}T^{\sigma_i\dagger}_{\alpha_i\alpha'_{i-1}}=\delta_{\alpha_{i-1}\alpha'_{i-1}}.
\end{equation}
Each tensor is then an isometry from $\ket{\alpha_{i-1}}$ to $\ket{\alpha_i,\sigma_i}$, which can be written as a unitary acting on an ancillary state $\ket{0}_i$,
\begin{equation}
T^{\sigma_i}_{\alpha_{i-1}\alpha_i}=\bra{\alpha_i,\sigma_i}U_i\ket{0_i,\alpha_{i-1}}.
\label{eq:isometry}
\end{equation}
The MPS is thereby realized as a sequential staircase of unitaries,
\begin{equation}
\ket\psi=U_NU_{N-1}\cdots U_1\ket0,
\end{equation}
as illustrated in Fig.~\ref{fig:mps_loading}. For $\chi_{\max}=2$ the $U_i$ are two-qubit gates; for $\chi_{\max}>2$ they act on $1+\lceil\log_2\chi_{\max}\rceil$ qubits and must be compiled into native two-qubit gates. The $\chi=2$ case of Fig.~\ref{fig:mps_loading} serves only as an illustration of the loading principle in the simplest nontrivial setting; it does not represent the bond dimensions of the benchmarks, which are listed in Table~\ref{tab:benchmarks}.

\section{Tensor-network simulation framework}
\label{sec:tn}
\begin{figure}[t]
\centering
\includegraphics[width=0.75\textwidth]{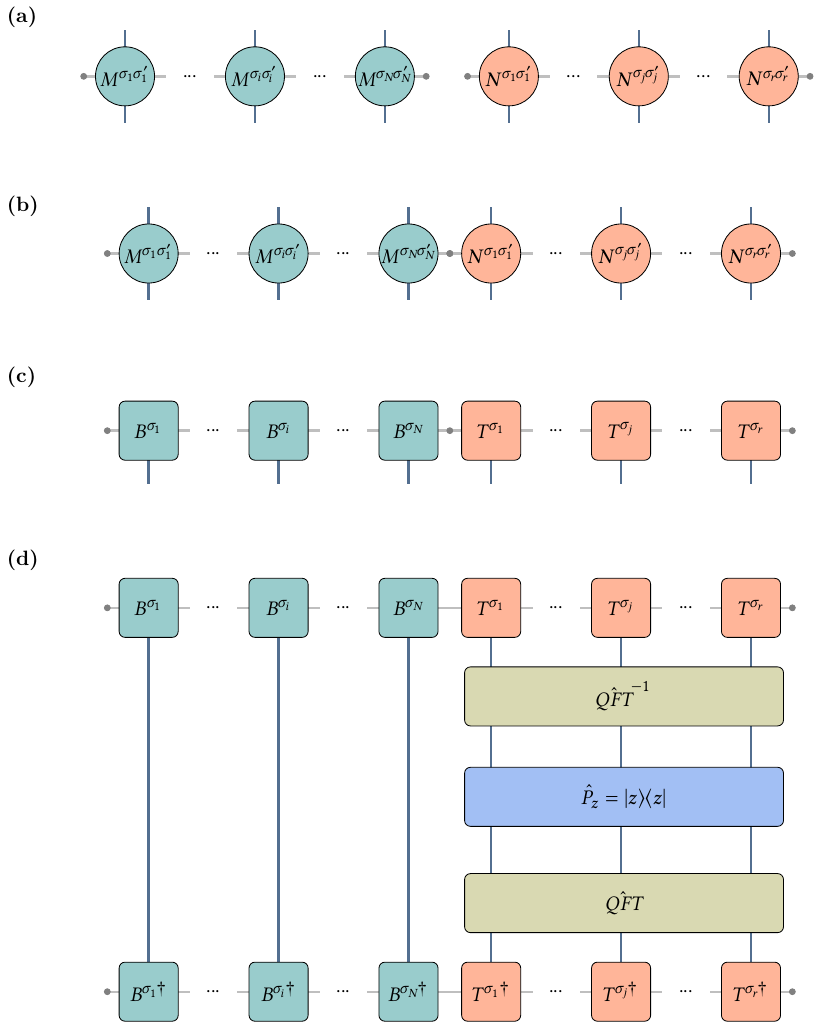}
\caption{Tensor-network representation of the circuit of Fig.~\ref{fig:circuit}. (a)~MPOs of the system Hamiltonian $\hat H$ (tensors $M^{\sigma_i\sigma_i'}$, $N$ sites, teal) and of the pointer momentum $\hat p$ (tensors $N^{\sigma_j\sigma_j'}$, $r$ sites, orange), each with open boundary conditions. (b)~The two MPOs attached to form the MPO of $\hat H\otimes\hat p$ on the composite chain of $N+r$ sites [Eq.~\eqref{eq:mpo_bond}]. (c)~Initial state: the DMRG-optimized system MPS (tensors $B^{\sigma_i}$) attached to the product-state MPS of the pointer (tensors $T^{\sigma_j}$, bond dimension 1). (d)~After TDVP evolution under $e^{-it\hat H\otimes\hat p}$ to time $t$, the pointer distribution is obtained as $P(z;t)=\la\Psi(t)|\,\hat I_{\mathrm{sys}}\otimes \mathrm{QFT}\,\hat P_z\,\mathrm{QFT}^{-1}|\Psi(t)\ra$ with $\hat P_z=\ket z\!\bra z$, by contracting the system tensors with their conjugates and applying the inverse QFT to the pointer [Eq.~\eqref{eq:readout_tn}].}
\label{fig:tn_scheme}
\end{figure}

This section describes the simulation framework, which is the first contribution of this paper. Its purpose is to benchmark the coupled system--pointer dynamics of the protocol in Sec.~\ref{sec:background} for a given DMRG input. The workflow is summarized in Fig.~\ref{fig:tn_scheme}.

\subsection{Matrix product states}
\label{sec:tn_mps}
An MPS of $L$ sites [Eq.~\eqref{eq:MPS}] is invariant under the gauge transformations $T^{\sigma_i}\to X_{i-1}^{-1}T^{\sigma_i}X_i$ on each bond, which we fix by canonical forms~\cite{schollwockDensitymatrixRenormalizationGroup2011}. A tensor $A^{\sigma_i}$ is left-canonical if $\sum_{\sigma_i}A^{\sigma_i\dagger}A^{\sigma_i}=I$ and a tensor $B^{\sigma_i}$ is right-canonical if $\sum_{\sigma_i}B^{\sigma_i}B^{\sigma_i\dagger}=I$. In the mixed canonical form with respect to bond $\ell$,
\begin{equation}
\ket\psi=\sum_{\alpha_\ell=1}^{\chi_\ell}\lambda^{[\ell]}_{\alpha_\ell}\,\ket{\Phi^L_{\alpha_\ell}}\otimes\ket{\Phi^R_{\alpha_\ell}},
\label{eq:schmidt}
\end{equation}
the left (right) block states $\ket{\Phi^{L}_{\alpha_\ell}}$ ($\ket{\Phi^{R}_{\alpha_\ell}}$), built from the $A$ ($B$) tensors, are orthonormal, and the $\lambda^{[\ell]}_{\alpha_\ell}$ are the Schmidt coefficients of the bipartition at bond $\ell$. The bipartite entanglement entropy is
\begin{equation}
S_\ell=-\sum_{\alpha_\ell}\big(\lambda^{[\ell]}_{\alpha_\ell}\big)^2\ln\big(\lambda^{[\ell]}_{\alpha_\ell}\big)^2\le\ln\chi_\ell ,
\label{eq:entropy}
\end{equation}
so the bond dimension needed to represent a state is set by its entanglement. Truncating the Schmidt spectrum to the $\chi$ largest values is the optimal rank-$\chi$ approximation in the 2-norm, with discarded weight $\varepsilon_\ell=\sum_{\alpha_\ell>\chi}(\lambda^{[\ell]}_{\alpha_\ell})^2$. Expectation values $\la\psi|\hat O|\psi\ra$ with an MPO $\hat O$ of bond dimension $D$ are contracted site by site through transfer tensors at a cost $\mathcal O(L\,d^2D\chi^3)$ for local dimension $d$~\cite{schollwockDensitymatrixRenormalizationGroup2011}; for the qubit Hamiltonians considered here $d=2$.

\subsection{MPO representation of $\hat H$}
\label{sec:tn_mpo}
To obtain the DMRG state and to evolve the coupled system, the Hamiltonian of Eq.~\eqref{eq:H_pauli} is represented as an MPO,
\begin{equation}
\hat H=\sum_{\substack{\sigma_1\ldots\sigma_N\\ \sigma'_1\ldots\sigma'_N}}M^{\sigma_1\sigma'_1}\cdots M^{\sigma_i\sigma'_i}\cdots M^{\sigma_N\sigma'_N}\,\ket{\sigma_1\cdots\sigma_N}\!\bra{\sigma'_1\cdots\sigma'_N},
\label{eq:MPO}
\end{equation}
where each $M^{\sigma_i\sigma'_i}$ is a $D_{i-1}\times D_i$ matrix with $D_0=D_N=1$ (open boundary conditions), $D_H=\max_iD_i$ is the MPO bond dimension, and $\ket{\sigma_i}$, $\ket{\sigma'_i}$ are local basis states. Equivalently, each $M^{[i]}$ is a $D_{i-1}\times D_i$ matrix whose entries are single-site operators in $\{\hat I,\hat X,\hat Y,\hat Z\}$ multiplied by numbers, and $\hat H$ is the matrix product $M^{[1]}M^{[2]}\cdots M^{[N]}$ with operator-valued entries multiplied by the tensor product.

\paragraph*{Finite-state-machine picture.} For regular, short-range Hamiltonians the operator-valued matrices can be written down directly as a finite state machine~\cite{crosswhite2008}: the MPO bond index labels the ``state'' of a partially placed term, and the matrix entry $M^{[i]}_{ab}$ is the operator placed on site $i$ in the transition $a\to b$. For example, a nearest-neighbor chain $\sum_i J\,(\hat X_i\hat X_{i+1}+\hat Y_i\hat Y_{i+1}+\hat Z_i\hat Z_{i+1})$ has the bulk tensor
\begin{equation}
M^{[i]}=
\begin{pmatrix}
\hat I & \hat X & \hat Y & \hat Z & 0\\
0 & 0 & 0 & 0 & J\hat X\\
0 & 0 & 0 & 0 & J\hat Y\\
0 & 0 & 0 & 0 & J\hat Z\\
0 & 0 & 0 & 0 & \hat I
\end{pmatrix},
\label{eq:fsm_heisenberg}
\end{equation}
with boundary vectors selecting the first row at site 1 and the last column at site $N$, so that $D=5$. State 1 means ``no operator placed yet'', states 2--4 ``a Pauli operator has been placed on the previous site'', and state 5 ``the term is complete''.

\paragraph*{Graph-based construction for general Pauli sums.} The periodic triangular lattice, mapped onto a chain, and in particular the Jordan--Wigner-transformed molecular Hamiltonians (Appendix~\ref{app:jw}) contain thousands of long-range Pauli strings, for which no simple regular pattern exists. For these we use the graph-based (matrix-product-diagram) construction~\cite{crosswhite2008, Keller2015MPO, Ren2020} as implemented in Ref.~\cite{javanmard2026coefficientdecoupledmatrixproductoperators}. Each Pauli string $\hat h_i$ is split at every bond $\ell$ into a left fragment on sites $1,\dots,\ell$ and a right fragment on sites $\ell+1,\dots,N$. The distinct left fragments form the vertices of a layered left graph, and the distinct right fragments those of a right graph; adjacent layers are connected by edges labeled with the single-site Pauli operator that extends a fragment by one site. Identical fragments shared by many terms are merged into a single vertex. The coefficients $c_i$ are attached to the edges where the left and right graphs meet. The MPO tensor $M^{[i]}$ is then the adjacency matrix between the vertex layers $i-1$ and $i$, with the edge labels as operator-valued entries, and $D_i$ equals the number of vertices in layer $i$. This deduplication keeps the bond dimension far below the number of terms $K$; Appendix~\ref{app:mpo_graph} (Fig.~\ref{fig:mpo_graph}) illustrates the construction for pyridine and $\mathrm{BH}_3$ and gives their MPO bond dimensions.

\paragraph*{Compression.} The MPO obtained from the graph is exact but not necessarily minimal. We compress it by bringing it into left-canonical form in a left-to-right sweep of QR decompositions, followed by a right-to-left sweep of singular value decompositions of the bond matrices, discarding singular values below a relative threshold $\epsilon=10^{-8}$~\cite{javanmard2026coefficientdecoupledmatrixproductoperators}. At this threshold the compression is numerically lossless: only exactly vanishing (or round-off) singular values are removed, so the compressed MPO represents the same operator with a smaller bond dimension. The resulting MPO bond dimensions agree with those of the independent \textsc{Renormalizer} construction~\cite{Ren2020} for molecular Hamiltonians of the size considered here~\cite{javanmard2026coefficientdecoupledmatrixproductoperators}.

\subsection{MPO of the pointer and of the coupling $\hat H\otimes\hat p$}
\label{sec:tn_mpo_coupling}
The pointer momentum, Eq.~\eqref{eq:pointer}, is a sum of single-site terms. Its MPO has bond dimension 2, with bulk tensor
\begin{equation}
N^{[j]}=
\begin{pmatrix}
\hat I & \hat p_j\\
0 & \hat I
\end{pmatrix},
\qquad \hat p_j=2^{-j}\,\frac{\hat I-\hat Z_j}{2},
\label{eq:pointer_mpo}
\end{equation}
left boundary vector $(1,0)$ and right boundary vector $(0,1)^T$: state 1 means ``$\hat p_j$ not yet placed'' and state 2 ``$\hat p_j$ placed''.

\begin{figure}[t]
\centering
\includegraphics[width=0.92\textwidth]{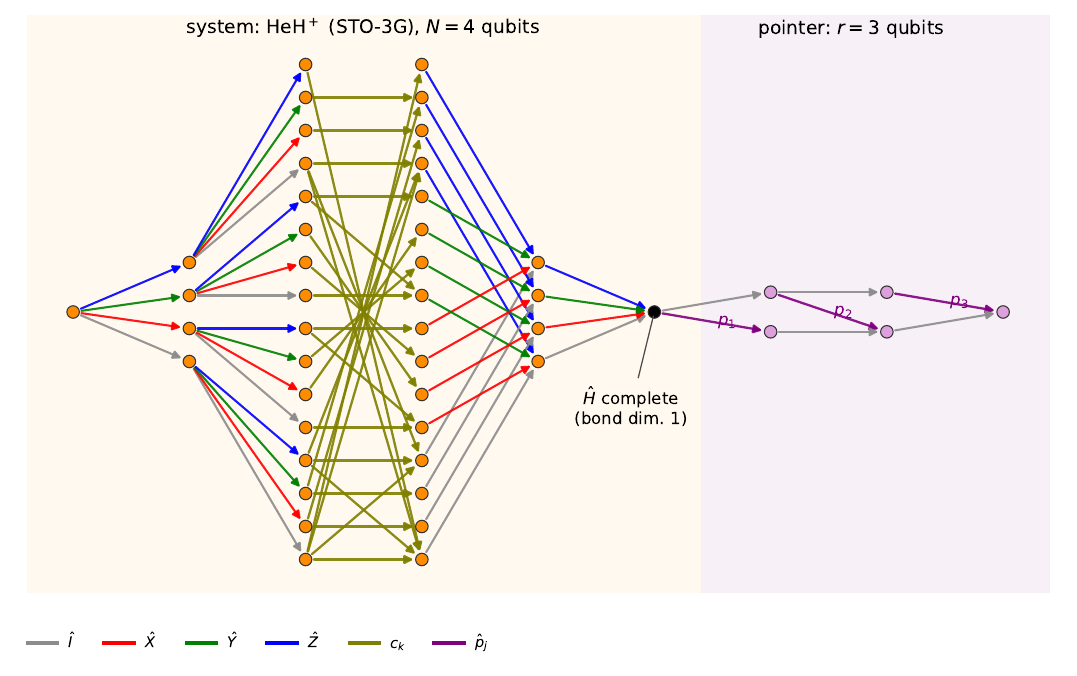}
\caption{Graph of the MPO of $\hat H\otimes\hat p$ for a molecule coupled to the pointer: $\mathrm{HeH}^+$ in the STO-3G basis (Jordan--Wigner, $N=4$ qubits, bond length $0.772$\,\AA, all 27 Pauli strings) followed by an $r=3$ pointer. Left (orange background): the molecular graph built as in Fig.~\ref{fig:mpo_graph}, i.e., left graph of Pauli prefixes, coefficient edges $c_k$ at the central cut, and right graph of suffixes; edge colors give the single-site operator. All terms end in a single vertex (black), at which the operator $\hat H$ is complete: the bond between the last system site and the first pointer site has MPO bond dimension 1. Right (purple background): the pointer MPO of Eq.~\eqref{eq:pointer_mpo} with two states per bond, ``$\hat p$ not yet placed'' (upper) and ``placed'' (lower); along each path exactly one $\hat p_j$ is placed, and the right boundary keeps only the ``placed'' state. The number of vertices per layer, $1,4,16,16,4,1,2,2,1$, is the bond dimension of each bond before compression. Conclusion: coupling the pointer appends a bond-dimension-2 chain to the molecular MPO, so $D_{H\otimes p}=\max(D_H,2)=D_H$ [Eq.~\eqref{eq:mpo_bond}].}
\label{fig:mpo_coupled}
\end{figure}

The coupling $\hat H\otimes\hat p$ acts on the composite chain of $N+r$ sites, with the pointer sites placed after the system sites [Fig.~\ref{fig:tn_scheme}(b)]. Its MPO is obtained by concatenating the two MPOs: the system tensors $M^{[1]},\dots,M^{[N]}$ contract to the operator $\hat H$ carried by the single outgoing channel at bond $N$ ($D_N=1$), which is then fed into the left boundary of the pointer MPO. Hence
\begin{equation}
\hat H\otimes\hat p=M^{[1]}\cdots M^{[N]}\;N^{[1]}\cdots N^{[r]},
\qquad
D_{H\otimes p}=\max(D_H,2)=D_H ,
\label{eq:mpo_bond}
\end{equation}
where the boundary vectors of Eq.~\eqref{eq:pointer_mpo} are understood. Figure~\ref{fig:mpo_coupled} shows this structure for a molecular Hamiltonian. Coupling the pointer therefore does not increase the MPO bond dimension, and the bond between the last system site and the first pointer site has MPO bond dimension 1. By contrast, building $\hat H\otimes\hat p$ directly as a sum of $rK$ Pauli strings, Eq.~\eqref{eq:coupling}, and compressing it would lead to the same operator at a higher construction cost.

\subsection{DMRG initial state}
\label{sec:tn_dmrg}
We employ DMRG to construct an MPS approximation of the ground state of $\hat H$ with bond dimension $\chidmrg$, $\ket{\psi_{\chidmrg}}=\hat U_{\mathrm{DMRG}}\ket{\psi_0}$, starting from an initial product state; for the molecular problems this is the Hartree--Fock state (Sec.~\ref{sec:molecules}). In each two-site DMRG step, the MPS is brought into mixed canonical form around sites $(i,i+1)$ and the two-site tensor $\Theta^{\sigma_i\sigma_{i+1}}$ is updated by the lowest eigenvector of the effective Hamiltonian
\begin{equation}
H^{\mathrm{eff}}_{i,i+1}\,\Theta=E\,\Theta,
\end{equation}
obtained by contracting the MPO with the left and right environments of the orthonormal block states; the eigenproblem is solved with the Lanczos method. The updated tensor is split by a singular value decomposition, truncated to at most $\chidmrg$ singular values, and the procedure is swept back and forth through the chain until the energy converges~\cite{whiteDensityMatrixFormulation1992, schollwockDensitymatrixRenormalizationGroup2011}. In a large-scale application, $\chidmrg$ would be the largest bond dimension affordable on a classical computer, and one expects $|\la\psi_{\chidmrg}|\Omega\ra|\to1$ as $\chidmrg$ increases. To test the protocol, we instead fix $\chidmrg$ to small values (Table~\ref{tab:benchmarks}), so that the initial energy $E_{\mathrm{init}}=\la\psi_{\chidmrg}|\hat H|\psi_{\chidmrg}\ra$ is far from $E_0$. For the spin model, DMRG starts from a random product state, discards singular values below $10^{-7}$, and stops when the relative change of the energy between sweeps falls below $10^{-6}$ or after at most six sweeps; the input energies therefore reflect both the bond dimension and this short optimization.

The pointer state $\ket{x=0}$ of Eq.~\eqref{eq:pointer_init} is the product state $\ket{+}^{\otimes r}$ with $\ket+=(\ket0+\ket1)/\sqrt2$, i.e., an MPS of bond dimension 1. Attaching it to the system MPS gives the initial state of the composite chain [Fig.~\ref{fig:tn_scheme}(c)],
\begin{equation}
\ket{\Psi(0)}=\ket{\psi_{\chidmrg}}\otimes\ket{+}^{\otimes r}.
\end{equation}

\subsection{TDVP evolution}
\label{sec:tn_tdvp}
The composite state is evolved under $\hat H\otimes\hat p$ with the TDVP~\cite{haegeman_time-dependent_2011, haegeman_unifying_2016} [Fig.~\ref{fig:tn_scheme}(d)], which projects the Schr\"odinger equation onto the tangent space of the MPS manifold and therefore conserves the norm and, for a time-independent generator, the energy $\la\hat H\otimes\hat p\ra$. All benchmarks of the main text use two-site TDVP throughout (Appendix~\ref{app:tdvp}), which allows the bond dimension to grow as entanglement builds up. The local exponentials are evaluated with a Krylov (Lanczos) method, and the two-site updates are truncated according to a singular-value tolerance of $10^{-6}$ and a maximal bond dimension; once the bond dimension has reached this maximum, two-site TDVP continues with truncation rather than switching to one-site TDVP. The maximal bond dimension is $\chitdvp^{\max}=128$ for the spin model and $\chitdvp^{\max}=100$ for the molecules, and the time step is $\delta t=0.05$ (in units of $1/J$ for the spin model and $1/\mathrm{Ha}$ for the molecules).

\subsection{Pointer readout from the MPS}
\label{sec:tn_readout}
At each time $t=n\,\delta t$, the pointer is read out by tracing out the system and applying the inverse QFT [Fig.~\ref{fig:tn_scheme}(d)]. Because the pointer occupies the last $r$ sites of the chain, this is inexpensive. Bringing $\ket{\Psi(t)}$ into mixed canonical form with respect to the system--pointer bond $N$, Eq.~\eqref{eq:schmidt}, the system block states are orthonormal and the partial trace over the system reduces to a sum over the bond index:
\begin{equation}
\rho_{\mathrm p}(t)=\Tr_{\mathrm{sys}}\ket{\Psi(t)}\!\bra{\Psi(t)}=\sum_{\alpha_N}\big(\lambda^{[N]}_{\alpha_N}\big)^2\,\ket{\Phi^R_{\alpha_N}}\!\bra{\Phi^R_{\alpha_N}} ,
\end{equation}
a $2^r\times2^r$ matrix ($16\times16$ for $r=4$) obtained by contracting only the $r$ pointer tensors. The readout distribution is then
\begin{equation}
P(x;t)=\la x|\,\mathrm{QFT}^{-1}\rho_{\mathrm p}(t)\,\mathrm{QFT}\,|x\ra ,
\label{eq:readout_tn}
\end{equation}
which is exact for the given MPS. The same Schmidt coefficients give the system--pointer entanglement entropy, Eq.~\eqref{eq:entropy} at bond $N$,
\begin{equation}
S(t)=-\Tr\rho_{\mathrm p}\ln\rho_{\mathrm p}=-\sum_{\alpha_N}\big(\lambda^{[N]}_{\alpha_N}\big)^2\ln\big(\lambda^{[N]}_{\alpha_N}\big)^2\le r\ln2 ,
\label{eq:S_pointer}
\end{equation}
where the bound follows from $\chi_N\le2^r$. We use $S(t)$ in Sec.~\ref{sec:entanglement} as a diagnostic of the quality of the initial state.

\paragraph*{Energy estimation.} For each outcome $x$, the time trace $P(x;t)$ is fitted by least squares to the single-eigenvalue form of Eq.~\eqref{eq:pointerDist}, $|f(E,x)|^2$ as a function of $t$, with the energy $E$ as the only free parameter. To avoid local minima of the least-squares cost, $E$ is first scanned on a fine grid in the window $[E_{\mathrm{init}}-1, E_{\mathrm{init}}]$ and then refined locally. The window is a prior that must contain the target energy; for all benchmarks the recovered energies are unchanged when it is widened to $[E_{\mathrm{init}}-3, E_{\mathrm{init}}]$, and no row estimate lies on its edge. This gives one estimate $E_{\mathrm{es}}(x)$ per outcome, i.e., $2^r=16$ nearly independent estimates of the same energy. We report their median as $E_{\mathrm{estimated}}$, which is robust against an occasional failed fit, and their spread as a measure of consistency. The estimate as a function of the evolution time (Fig.~\ref{fig:convergence}) is obtained by fitting only the data up to that time.

\paragraph*{Implementation.} All tensor-network calculations (MPO construction and compression, DMRG, TDVP, and pointer readout) were performed with an in-house Python implementation, the same code base as in Ref.~\cite{javanmard2026coefficientdecoupledmatrixproductoperators}. Molecular integrals, active spaces, and FCI reference energies were obtained with PySCF~\cite{sun_python-based_2017} through Tangelo~\cite{tangelo}, and the Jordan--Wigner mapping with OpenFermion~\cite{McClean2017}.

\subsection{Scope of the simulation}
\label{sec:tn_scope}
The TDVP approximates the exact unitary $e^{-it\hat H\otimes\hat p}$ within the MPS manifold. It does not decompose $\hat H\otimes\hat p$ into Pauli-string exponentials; its own time-step error arises from the symmetric splitting of the tangent-space projector into local updates and is unrelated to the product-formula error of a compiled circuit. Its error is controlled by $\delta t$, $\chitdvp^{\max}$, and the growth of the bond dimension from the initial state (see the validation below). The benchmarks therefore describe the ideal protocol: they contain no product-formula (Trotter) error of a circuit implementation, no hardware noise, and no shot noise; the effect of a finite number of measurements is quantified in Appendix~\ref{app:sampling}. Where our figures show a Trotter-step axis $n=t/\delta t$, it indicates the number of steps a first-order product formula with the same nominal step size would require; it serves as a reference for the auxiliary implementation model of Sec.~\ref{sec:resources} and is not a simulation of Trotterized circuit dynamics. Product-formula error is discussed separately in Appendix~\ref{app:trotter_error_analysis}.

\paragraph*{Validation against exact evolution.} We validated the complete pipeline---MPO of $\hat H\otimes\hat p$, DMRG input, TDVP evolution, and pointer readout---against an exact state-vector simulation of the ideal circuit for a system small enough to be diagonalized: the Heisenberg model on a periodic $2\times3$ triangular lattice (6 spins) with an $r=3$ pointer, starting from a $\chidmrg=3$ DMRG state. For $\delta t=0.05$ the TDVP distribution deviates from the exact one by at most $\max_{x,t}|P_{\mathrm{TN}}(x;t)-P_{\mathrm{exact}}(x;t)|=1.0\times10^{-2}$ over $t\le1$, and the deviation decreases with the time step ($5.2\times10^{-3}$ at $\delta t=0.025$, $2.6\times10^{-3}$ at $\delta t=0.0125$). An independent implementation of two-site TDVP in the inverse canonical gauge~\cite{Secular2020ParallelTDVP} reproduces these distributions to $1.5\times10^{-6}$. With the bond dimension kept sufficiently large, the discarded weight and the drift of $\la\hat H\otimes\hat p\ra$ are below $10^{-13}$ in this test. The remaining deviation is not a time-step error of the integrator, which converges much faster: it is the projection error of two-site TDVP while the bond dimension grows from the small value of the initial product state $\ket{\psi_{\chidmrg}}\otimes\ket{+}^{\otimes r}$. Because $\hat H\otimes\hat p$ couples every system term to the pointer, and the periodic lattice maps to long-range couplings along the chain, the two-site tangent space does not contain the full action of the generator until the bond dimension has grown, which takes a number of steps; a smaller $\delta t$ reduces the time this takes, hence the linear decrease. If the bond dimension is first saturated by a few steps of negligible duration, the deviation drops to $2\times10^{-5}$, independently of $\delta t$. On the fitted energy, this growth error amounts to $8\times10^{-3}$ in this test (estimate over $t\le8$), compared with $2\times10^{-3}$ after saturating the bond dimension. All benchmarks of Sec.~\ref{sec:benchmarks} start from the unsaturated product state and therefore contain an error of this type; for the Heisenberg model it is of the same order as the deviation of the recovered energy from $E_0$ (Sec.~\ref{sec:heisenberg}). Global subspace expansion or an initial bond-dimension saturation removes it at negligible cost and is recommended for future applications.

\section{Benchmarks}
\label{sec:benchmarks}
We consider two classes of strongly correlated systems: quantum spin systems and electronic-structure problems. Table~\ref{tab:benchmarks} collects the parameters and results. In all cases the pointer has $r=4$ qubits, and the reference energy $E_{\mathrm{ref}}$ is the exact ground-state energy of the same qubit Hamiltonian (exact diagonalization for the spin model, FCI within the chosen orbital space for the molecules). We deliberately choose system sizes for which the exact ground state is known, so that the recovered energies can be assessed; the framework itself applies to larger systems.

\begin{table}[t]
\caption{Benchmark parameters and results. $N$: number of system qubits; $r$: number of pointer qubits; $\chidmrg$: bond dimension of the DMRG initial state; $E_{\mathrm{init}}$: its variational energy; $t_{\max}$: longest evolution time simulated; $E_{\mathrm{estimated}}$: energy recovered from the pointer; $E_{\mathrm{ref}}$: exact reference energy; $\Delta E=|E_{\mathrm{estimated}}-E_{\mathrm{ref}}|$. TDVP parameters: $\delta t=0.05$; $\chitdvp^{\max}=128$ (Heisenberg) and $100$ (molecules). The molecular Hamiltonians of $\mathrm{H}_8$ and pyridine are $\mathbb Z_2$-tapered from 16 to 13 qubits. $E_{\mathrm{estimated}}$ is the median of the $2^r=16$ row estimates $E_{\mathrm{es}}(x)$. Energies in units of $J$ (Heisenberg model) or Hartree (molecules). Chemical accuracy is $1.59\times10^{-3}$\,Ha. (8e,8o) denotes an active space of 8 electrons in 8 spatial orbitals.}
\label{tab:benchmarks}
\begin{ruledtabular}
\begin{tabular}{lccccccccc}
System & $N$ & $r$ & $\chidmrg$ & $E_{\mathrm{init}}$ & $t_{\max}$ & $E_{\mathrm{estimated}}$ & $E_{\mathrm{ref}}$ & $\Delta E$ & Fig. \\
\hline
Heisenberg, $4\times3$ triangular, PBC & 12 & 4 & 20 & $-26.7333$ & 10 & $-27.7110$ & $-27.7232$ & $1.2\times10^{-2}$ & \ref{fig:heisenberg} \\
$\mathrm{H}_8$, STO-3G, FCI & 13 & 4 & 4 & $-4.1851$ & 30 & $-4.2708$ & $-4.2718$ & $1.1\times10^{-3}$ & \ref{fig:H8} \\
Pyridine, STO-3G, (8e,8o) active space & 13 & 4 & 2 & $-243.6607$ & 30 & $-243.6954$ & $-243.6968$ & $1.4\times10^{-3}$ & \ref{fig:pyridine} \\
$\mathrm{BH}_3$, STO-3G, FCI$^{a}$ & 16 & 4 & 6 & $-32.4852$ & 30 & $-32.5043$ & $-32.5050$ & $6.7\times10^{-4}$ & \ref{fig:BH3} \\
\end{tabular}
\end{ruledtabular}
\begin{flushleft}\footnotesize $^{a}$\,Electronic energies without the nuclear repulsion ($6.7353$\,Ha); TDVP with $\chitdvp^{\max}=300$ (Appendix~\ref{app:BH3}).\end{flushleft}
\end{table}

\subsection{Quantum spin systems: Heisenberg model on a triangular lattice}
\label{sec:heisenberg}
We consider interacting spins governed by the antiferromagnetic Heisenberg Hamiltonian
\begin{equation}
\hat H=\sum_{\la i,j\ra}J\big(\hat X_i\hat X_j+\hat Y_i\hat Y_j+\hat Z_i\hat Z_j\big),
\label{eq:heisenberg}
\end{equation}
where $J=1$ is the coupling strength, $\hat X_i$, $\hat Y_i$, $\hat Z_i$ are Pauli matrices acting on spin $i$, and $\la i,j\ra$ denotes nearest neighbors, on a $4\times3$ triangular lattice (12 qubits) with periodic boundary conditions. Ground-state simulations of such systems in more than one dimension are challenging because of long-range correlations and geometric frustration, and the two-dimensional periodic geometry, mapped onto a one-dimensional MPS, makes this a difficult case for DMRG at small bond dimension. The exact ground-state energy is $E_0=-27.7232\,J$. We use a DMRG initial state with $\chidmrg=20$, whose energy $E_{\mathrm{init}}=-26.7333\,J$ lies $0.99\,J$ above $E_0$.

\begin{figure}[t]
\centering
\includegraphics[width=0.72\textwidth]{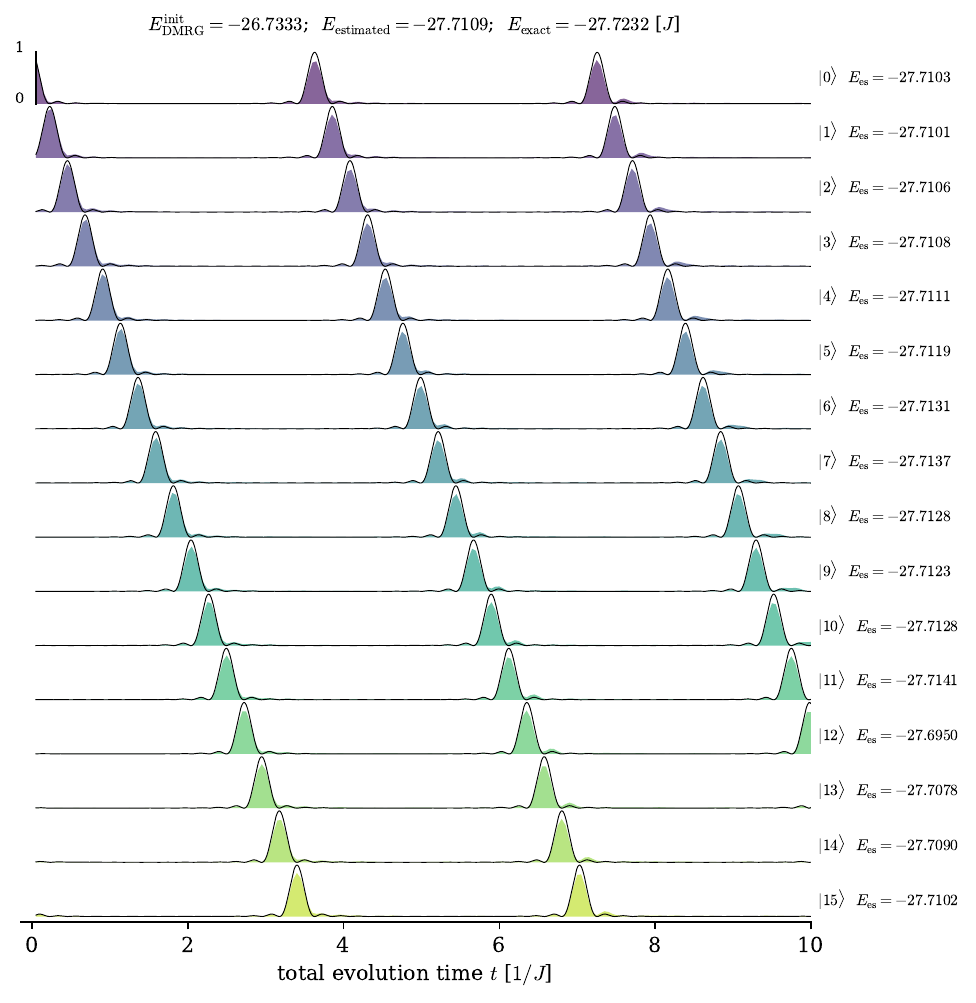}
\caption{Time-resolved pointer distribution for the antiferromagnetic Heisenberg model, Eq.~\eqref{eq:heisenberg}, on a $4\times3$ triangular lattice with periodic boundary conditions ($N=12$, $J=1$). Each row shows $P(x;t)$, the probability of pointer outcome $\ket{x}$ ($x=0,\dots,15$; $r=4$), as a function of the evolution time $t$ (horizontal axis, in units of $1/J$). Rows are offset vertically, $\ket{0}$ at the top; the scale at the top left indicates $P$ from 0 to 1. Filled areas: TDVP data; black lines: one-parameter fit of each row to Eq.~\eqref{eq:pointerDist}; right: the resulting estimate $E_{\mathrm{es}}(x)$. Header: DMRG initial-state energy, recovered energy $E_{\mathrm{estimated}}$ (median of the rows), and exact energy. Initial state: DMRG MPS with $\chidmrg=20$. TDVP: $\delta t=0.05/J$, $\chitdvp^{\max}=128$. The peak advances by one outcome every $2\pi/|E_0|\approx0.23/J$ and revives every $T_{\mathrm{rev}}\approx3.63/J$ [Eq.~\eqref{eq:revival}]. Conclusion: starting from an input $0.99\,J$ above the ground state, the pointer returns $E_0$ with a relative error of $4.4\times10^{-4}$.}
\label{fig:heisenberg}
\end{figure}

Figure~\ref{fig:heisenberg} shows the resulting time-resolved pointer distribution. The peaks follow the pattern of Sec.~\ref{sec:readout}: each row repeats with revival time $T_{\mathrm{rev}}=2^4\cdot2\pi/|E_0|\approx3.63/J$, and consecutive rows are shifted by $2\pi/|E_0|\approx0.23/J$. The peaks are sharp and nearly background-free, showing that the pointer resolves a single dominant energy component although the energy of the input is off by 3.6\%. The sixteen row estimates lie between $-27.6950\,J$ and $-27.7141\,J$, and their median $E_{\mathrm{estimated}}=-27.7110\,J$ differs from $E_0$ by $1.2\times10^{-2}\,J$ (relative error $4.4\times10^{-4}$), almost two orders of magnitude closer than the DMRG input. To separate the simulation error from that of the protocol, we repeated the run with the same DMRG protocol ($E_{\mathrm{init}}=-26.7325\,J$) twice, once as above and once with the bond dimension saturated before the evolution, which removes the TDVP error caused by the growth of the bond dimension from the initial product state (Sec.~\ref{sec:tn_scope}). The recovered energies are $-27.7130\,J$ and $-27.7171\,J$, i.e.\ $1.0\times10^{-2}\,J$ and $6.1\times10^{-3}\,J$ from $E_0$. The bond-growth error thus accounts for about $4\times10^{-3}\,J$ of the deviation. The remainder includes the truncation at $\chitdvp^{\max}=128$, which is significant for this two-dimensional lattice (accumulated discarded weight $1.1\times10^{-2}$, drift of $\la\hat H\otimes\hat p\ra$ of $8\times10^{-2}\,J$ at $t=10/J$), so the deviation of Fig.~\ref{fig:heisenberg} contains a substantial simulation contribution and does not isolate the error of the protocol itself. The role of the initial-state quality for this model is analyzed in Sec.~\ref{sec:entanglement}.

\subsection{Electronic-structure problems}
\label{sec:molecules}
The electronic Hamiltonian of a molecule with $N_e$ electrons in first quantization and atomic units reads
\begin{equation}
\hat H_e=\hat T_e+\hat V_{\mathrm{ext}}+\hat V_{ee}
=-\sum_i\frac{\nabla^2_{\mathbf r_i}}{2}-\sum_{i,j}\frac{Z_i}{|\mathbf R_i-\mathbf r_j|}+\sum_{i,j>i}\frac{1}{|\mathbf r_i-\mathbf r_j|},
\end{equation}
where $\mathbf R_i$ and $Z_i$ are the positions and charges of the nuclei and $\mathbf r_i$ are the electron positions. In second quantization,
\begin{equation}
\hat H=\sum_{kl}h_{kl}\,f_k^\dagger f_l+\sum_{klmn}h_{klmn}\,f_k^\dagger f_l^\dagger f_m f_n ,
\label{eq:electronic}
\end{equation}
where $f^\dagger_k$ and $f_k$ create and annihilate an electron in spin-orbital $k$, and the scalar coefficients are the one- and two-electron integrals over spin-orbitals $\phi_k(\mathbf x)$ with $\mathbf x=(\mathbf r,s)$,
\begin{equation}
h_{kl}=\int\phi^*_k(\mathbf x)\Big(-\frac{\nabla^2_{\mathbf r}}{2}-\sum_i\frac{Z_i}{|\mathbf R_i-\mathbf r|}\Big)\phi_l(\mathbf x)\,d\mathbf x,\qquad
h_{klmn}=\int\frac{\phi^*_k(\mathbf x_1)\phi^*_l(\mathbf x_2)\phi_m(\mathbf x_2)\phi_n(\mathbf x_1)}{|\mathbf r_1-\mathbf r_2|}\,d\mathbf x_1d\mathbf x_2
\end{equation}
\cite{bauer_quantum_2020, head-marsden_quantum_2021, tangelo, sun_python-based_2017, mcquarrie_quantum_2016, engel_quantum_2009, szabo_modern_1996, sajjan_quantum_2022, QuantumCompChemsitry, Cao2019, lanyon_towards_2010, whitfield_simulation_2011, PhysRevX.6.031007, collaborators_hartree-fock_2020, PhysRevX.8.031022, McClean2017}. The reported total energies also include the nuclear repulsion and, where applicable, the energy of frozen orbitals.

To implement Eq.~\eqref{eq:electronic} on a quantum computer, the fermionic Fock space is mapped to the qubit Hilbert space, so that the fermionic operators become qubit operators. The three standard mappings are the Jordan--Wigner (JW), Bravyi--Kitaev (BK), and parity transformations~\cite{QuantumCompChemsitry, jordan_uber_1928, Cao2019, whitfield_simulation_2011, seeley_bravyi-kitaev_2012}; we use the JW mapping (Appendix~\ref{app:jw}), which gives one qubit per spin-orbital and a Hamiltonian of the form of Eq.~\eqref{eq:H_pauli}. The resulting Pauli sums contain thousands of long-range strings; their MPOs are built with the graph-based construction and lossless compression developed and validated for molecular Hamiltonians in Ref.~\cite{javanmard2026coefficientdecoupledmatrixproductoperators} (Sec.~\ref{sec:tn_mpo}).

In molecular simulations, the orbitals are divided into core orbitals (always doubly occupied), active orbitals (occupied by zero, one, or two electrons), and external or virtual orbitals (unoccupied). Depending on the molecule and the target accuracy, the qubit requirement can be reduced by selecting an appropriate active space, commonly by retaining the frontier orbitals around the HOMO and LUMO levels; other standard approximations include the frozen-core approximation and truncation of virtual orbitals, e.g.\ by frozen natural orbitals~\cite{tangelo, sun_python-based_2017, mcquarrie_quantum_2016, engel_quantum_2009, szabo_modern_1996, verma_scaling_2021}. We consider two representative molecules in the main text: octahydrogen ($\mathrm{H}_8$), treated at the full-configuration-interaction (FCI) level without frozen orbitals, and pyridine ($\mathrm{C_5H_5N}$), treated in a selected active space. A third example, $\mathrm{BH}_3$, is given in Appendix~\ref{app:BH3}.

For each molecule we first express the Hartree--Fock reference $\ket{\Phi_0}$ as a product state and use it as the starting point of DMRG, obtaining $\ket{\psi_{\chidmrg}}=\hat U_{\mathrm{DMRG}}\ket{\Phi_0}$. In realistic applications to larger molecules, even DMRG with a large bond dimension may provide only an approximation of the ground state, and the remaining refinement would be performed on the quantum processor; here we fix $\chidmrg$ to small values to test this refinement. Molecular geometries are given in Appendix~\ref{app:geometry}.

\subsubsection{Octahydrogen: FCI without frozen orbitals}
\label{sec:H8}
$\mathrm{H}_8$ in the STO-3G basis (Slater-type orbitals, each represented by three primitive Gaussians) has 8 spatial orbitals, i.e., 16 spin-orbitals and 16 qubits after the Jordan--Wigner mapping; three $\mathbb Z_2$ symmetries (particle-number and spin parities) are removed by qubit tapering, leaving 13 qubits. It is treated at the FCI level without frozen orbitals. The FCI energy is $E_{\mathrm{FCI}}=-4.27182$\,Ha. The DMRG initial state with $\chidmrg=4$ has $E_{\mathrm{init}}=-4.1851$\,Ha, i.e., $87$\,mHa ($54$\,kcal/mol) above $E_{\mathrm{FCI}}$.

\begin{figure}[t]
\centering
\includegraphics[width=0.72\textwidth]{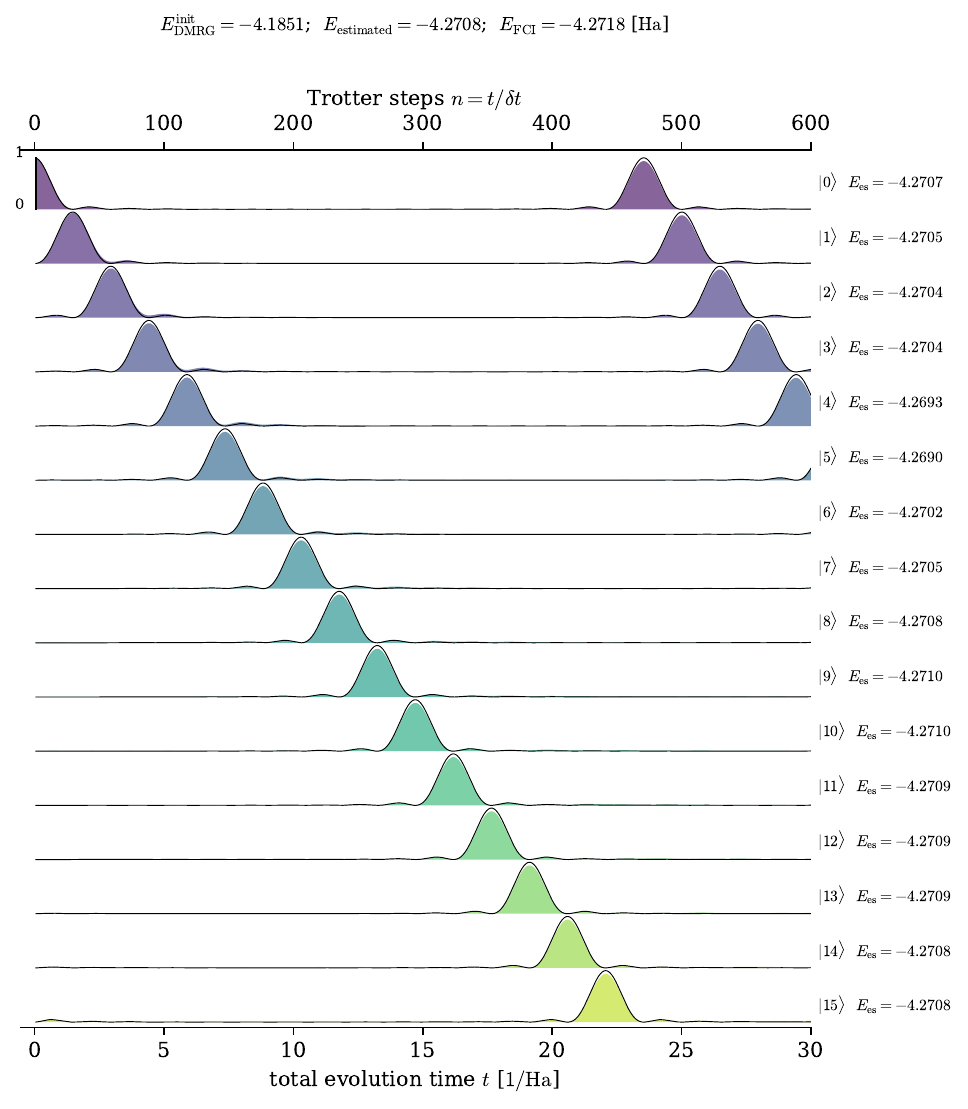}
\caption{Ground-state energy estimation for octahydrogen ($\mathrm{H}_8$), STO-3G, FCI ($N=13$ tapered qubits, $r=4$), geometry in Appendix~\ref{app:geometry}. Time-resolved pointer distribution; each row shows $P(x;t)$ for outcome $\ket{x}$ as a function of the total evolution time $t$ (bottom axis, $\mathrm{Ha}^{-1}$); the top axis gives the corresponding number of first-order Trotter steps $n=t/\delta t$ (reference for the auxiliary implementation model only; Sec.~\ref{sec:tn_scope}). Filled areas: TDVP data; black lines: one-parameter fit of each row; right: estimate $E_{\mathrm{es}}(x)$ from each row; header: $E_{\mathrm{estimated}}$ is the median of the rows. Initial state: DMRG MPS with $\chidmrg=4$, $E^{\mathrm{init}}_{\mathrm{DMRG}}=-4.1851$\,Ha. TDVP: $\delta t=0.05\,\mathrm{Ha}^{-1}$, $\chitdvp^{\max}=100$. All energies in Hartree. Conclusion: from an input 87\,mHa above the FCI energy, all sixteen rows agree to $2$\,mHa and the recovered energy lies $1.05$\,mHa from $E_{\mathrm{FCI}}$.}
\label{fig:H8}
\end{figure}

Figure~\ref{fig:H8} shows the time-resolved pointer distribution up to $t=30\,\mathrm{Ha}^{-1}$. The revival time is $T_{\mathrm{rev}}\approx23.5\,\mathrm{Ha}^{-1}$, so the window covers slightly more than one revival. All sixteen rows give estimates between $-4.2690$ and $-4.2710$\,Ha, and their median $E_{\mathrm{estimated}}=-4.2708$\,Ha lies $1.05$\,mHa from $E_{\mathrm{FCI}}$, within chemical accuracy ($1.59$\,mHa). Figure~\ref{fig:convergence} shows how the estimate converges when only the data up to time $t$ are fitted: the error decreases from $86$\,mHa at $t=0.5$ to $6$\,mHa at $t=5$, $2.4$\,mHa at $t=10$, and $1.05$\,mHa at $t\ge25\,\mathrm{Ha}^{-1}$, i.e., chemical accuracy is reached once the window covers about one revival ($n\approx500$ steps of $\delta t=0.05$). Increasing $t$ improves the spectral resolution of the pointer readout and reduces the fitted energy error, consistent with Sec.~\ref{sec:precision}; finite pointer discretization can cause small non-monotonic deviations at individual time points. The observed times are far shorter than the conservative single-shot bound, which for chemical accuracy at 90\% single-shot success requires $t\gtrsim2.4\times10^4\,\mathrm{Ha}^{-1}$ (Appendix~\ref{app:vN_chem}), because the energy is inferred from the positions of many sharp peaks rather than from a single outcome.

Within the tensor-network benchmark, the protocol combined with a pre-trained DMRG/MPS input thus reaches chemical accuracy for this instance. This should not be read as direct evidence that an explicit Suzuki--Trotter circuit would attain the same accuracy at the same nominal $\delta t$, since the product-formula error is not included (Sec.~\ref{sec:tn_scope}).

\subsubsection{Pyridine: FCI in a selected active space}
\label{sec:pyridine}
Pyridine ($\mathrm{C_5H_5N}$) is a heteroaromatic molecule and a standard benchmark in electronic-structure calculations. In the STO-3G basis we select an active space of the 3 orbitals below the HOMO and the 3 orbitals above the LUMO, together with the HOMO and LUMO. This reduces the problem from 29 molecular orbitals and 30 electrons to 8 active electrons in 8 active spatial orbitals, an active space denoted (8e,8o) (the remaining 17 lower orbitals are kept doubly occupied and the higher ones empty), corresponding to 16 qubits after the Jordan--Wigner mapping and 13 qubits after $\mathbb Z_2$ tapering. The exact ground-state energy in this active space is $E_{\mathrm{ref}}=-243.6968$\,Ha. We use a DMRG initial state with $\chidmrg=2$ and $E_{\mathrm{init}}=-243.6607$\,Ha, i.e., $36$\,mHa above $E_{\mathrm{ref}}$.

\begin{figure}[t]
\centering
\includegraphics[width=0.72\textwidth]{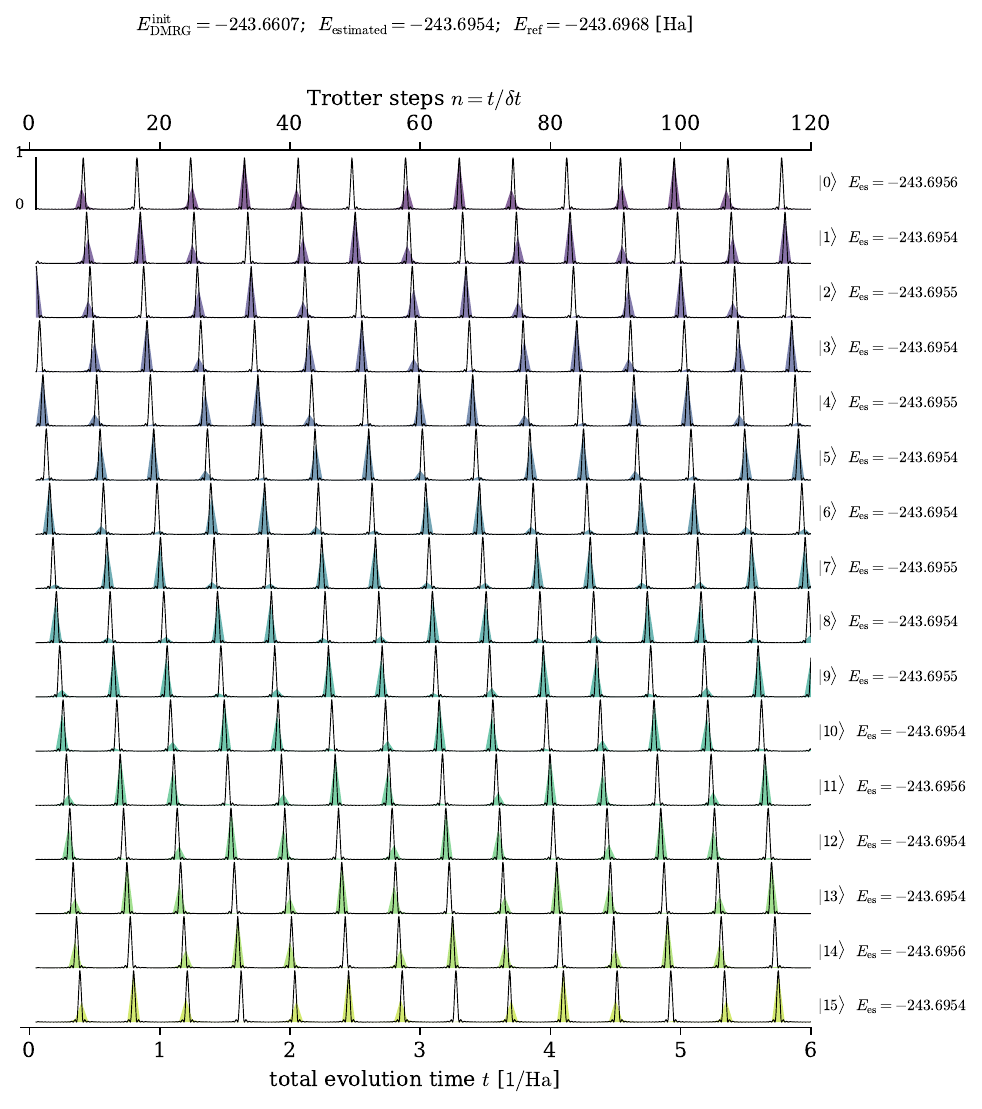}
\caption{Pyridine ($\mathrm{C_5H_5N}$), STO-3G, (8e,8o) active space ($N=13$ tapered qubits, $r=4$). Each row shows $P(x;t)$ for pointer outcome $\ket{x}$ as a function of the total evolution time $t$ (bottom axis, $\mathrm{Ha}^{-1}$); the top axis gives the corresponding number of Trotter steps $n=t/\delta t$ (reference only). Only the first $6\,\mathrm{Ha}^{-1}$ of the simulated window $t\le30\,\mathrm{Ha}^{-1}$ are displayed. Filled areas: TDVP data; black lines: one-parameter fit of each row; right: estimate $E_{\mathrm{es}}(x)$ from each row, fitted over the full window; header: median of the rows. Initial state: DMRG MPS with $\chidmrg=2$, $E^{\mathrm{init}}_{\mathrm{DMRG}}=-243.6607$\,Ha. TDVP: $\delta t=0.05\,\mathrm{Ha}^{-1}$, $\chitdvp^{\max}=100$. The short revival time $T_{\mathrm{rev}}\approx0.41\,\mathrm{Ha}^{-1}$ results from the large absolute energy. Peak heights vary because the peaks are narrow compared with $\delta t$; peak positions are unaffected. Conclusion: from a $\chidmrg=2$ input 36\,mHa above the active-space ground state, all sixteen rows agree to $0.2$\,mHa and the recovered energy lies $1.35$\,mHa from $E_{\mathrm{ref}}=-243.6968$\,Ha.}
\label{fig:pyridine}
\end{figure}

\begin{figure}[t]
\centering
\includegraphics[width=0.8\textwidth]{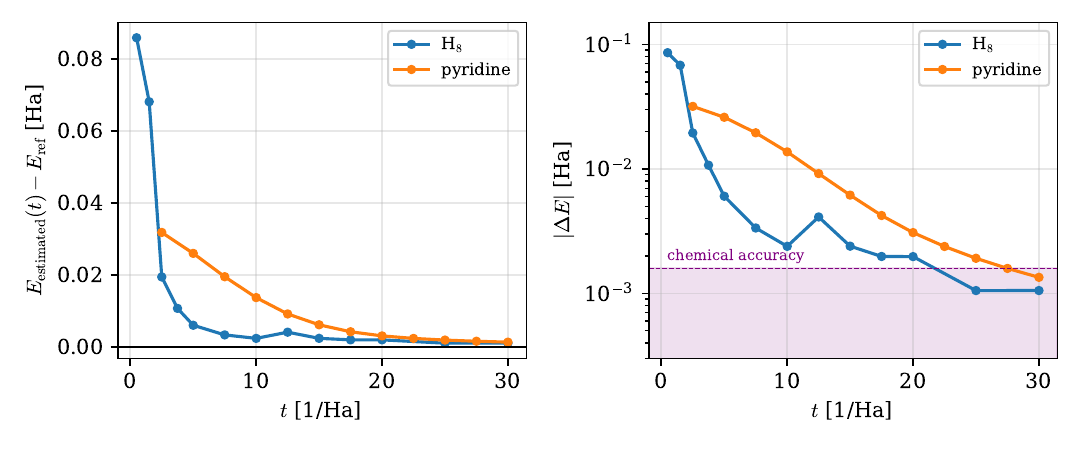}
\caption{Convergence of the recovered energy with the length of the fitted time window for $\mathrm{H}_8$ (blue) and pyridine (orange). At each $t$, every row of Figs.~\ref{fig:H8} and \ref{fig:pyridine} is fitted using only the data up to $t$, and $E_{\mathrm{estimated}}(t)$ is the median of the sixteen row estimates. Left: $E_{\mathrm{estimated}}(t)-E_{\mathrm{ref}}$ (Ha). Right: $|\Delta E|$ on a logarithmic scale; the shaded region marks chemical accuracy, $1\,\mathrm{kcal/mol}\approx1.59\times10^{-3}$\,Ha. Conclusion: both estimates approach the reference monotonically up to small fluctuations from the finite time sampling, and reach chemical accuracy at $t\approx25$ ($\mathrm{H}_8$) and $t\approx27.5\,\mathrm{Ha}^{-1}$ (pyridine).}
\label{fig:convergence}
\end{figure}

The coupled evolution was simulated up to $t=30\,\mathrm{Ha}^{-1}$; Fig.~\ref{fig:pyridine} displays the first $6\,\mathrm{Ha}^{-1}$ of the time-resolved pointer distribution. Because $|E_0|$ is large---it includes the nuclear-repulsion and core contributions---the peak spacing $2\pi/|E_0|\approx0.026\,\mathrm{Ha}^{-1}$ and the revival time $T_{\mathrm{rev}}\approx0.41\,\mathrm{Ha}^{-1}$ are short, so each row contains about fifteen peaks in the displayed window. The peaks are narrow compared with the time step $\delta t=0.05$, which is why their heights vary from peak to peak; their positions, and hence the fitted energies, are not affected. The variations of peak sharpness and contrast should thus not be interpreted as an optimization trajectory: each row is obtained from the same pre-trained initial state, and the structure reflects phase accumulation together with the finite pointer discretization and the modulo-$2^r$ readout. Fitting the full window $t\le30\,\mathrm{Ha}^{-1}$, all sixteen rows give estimates between $-243.6954$ and $-243.6956$\,Ha, and their median $E_{\mathrm{estimated}}=-243.6954$\,Ha lies $1.35$\,mHa ($0.85$\,kcal/mol) from $E_{\mathrm{ref}}$, within chemical accuracy. The estimate converges with the length of the fitted window (Fig.~\ref{fig:convergence}): using data up to $t=10$, $20$, and $27.5\,\mathrm{Ha}^{-1}$, the error is $13.7$, $3.1$, and $1.6$\,mHa, respectively, so chemical accuracy is reached at $t\approx27.5\,\mathrm{Ha}^{-1}$. Increasing $t$ thus improves the spectral resolution and stabilizes the estimate, as for $\mathrm{H}_8$. Subtracting a known constant from $\hat H$ before coupling it to the pointer would lengthen $\Delta t$ and $T_{\mathrm{rev}}$; such a shift is also the standard way to satisfy the no-aliasing condition, Eq.~\eqref{eq:no_aliasing}.

\subsection{Initial-state quality and system--pointer entanglement}
\label{sec:entanglement}
Equation~\eqref{eq:Pxt} shows that the input enters the readout only through its weights $|a_j|^2$ on the eigenstates. For an exact eigenstate, the evolved state remains a product $\ket{\psi_j}\otimes(\text{pointer state})$ and the system--pointer entanglement vanishes; every additional eigencomponent correlates the pointer position with a different energy and generates entanglement. The entanglement entropy $S(t)$ of the pointer (Sec.~\ref{sec:tn_readout}) is therefore a direct diagnostic of the quality of the initial state, and it also determines the bond dimension that the TDVP simulation requires.

\begin{figure}[t]
\centering
\includegraphics[width=0.55\textwidth]{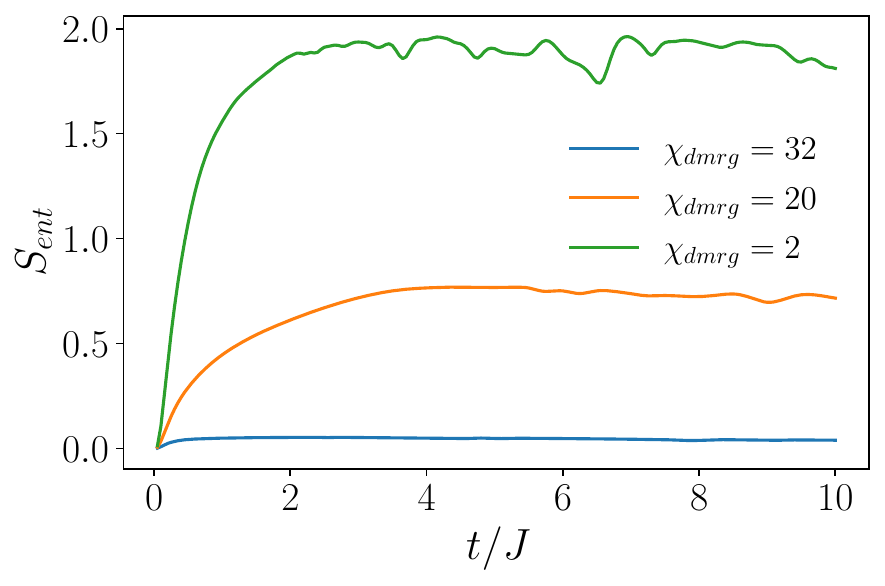}
\caption{Entanglement entropy $S(t)$ of Eq.~\eqref{eq:S_pointer} (axis label $S_{\mathrm{ent}}$) between the $r=4$ pointer (ancilla) qubits and the system during the evolution under $\hat H\otimes\hat p$, computed from the eigenvalues of the reduced density matrix of the pointer, obtained by tracing out the system (natural logarithm; upper bound $r\ln2\approx2.77$), for the Heisenberg model on the $4\times3$ triangular lattice with periodic boundary conditions (same setup as Fig.~\ref{fig:heisenberg}). Curves: DMRG initial states with $\chidmrg=32$ (blue), $20$ (orange), and $2$ (green). Horizontal axis: evolution time in units of $1/J$. TDVP: $\delta t=0.05/J$, $\chitdvp^{\max}=128$. Conclusion: the more accurate the DMRG initial state, the less entanglement the measurement generates; for $\chidmrg=32$ it is almost zero, i.e., the input is nearly an eigenstate of $\hat H$.}
\label{fig:entanglement}
\end{figure}

Figure~\ref{fig:entanglement} shows $S(t)$ for the Heisenberg model of Sec.~\ref{sec:heisenberg} with three DMRG initial states. For $\chidmrg=32$, the initial state is already very close to the ground state and the entropy stays below $\approx0.05$ throughout the evolution, with only weak growth. For $\chidmrg=20$, the input of Fig.~\ref{fig:heisenberg}, $S$ grows to $\approx0.75$ by $t\approx4/J$ and then saturates. For $\chidmrg=2$, where the overlap with the ground state is smaller, $S$ grows to $\approx1.9$ within $t\approx2/J$ and saturates with fluctuations. A better initial overlap thus leads to weaker dynamical mixing and less system--pointer entanglement.

For a high-quality DMRG/MPS initial state, the von Neumann measurement dynamics does not strongly delocalize the state over many energy components. Consequently, the growth of system--pointer entanglement remains moderate and the TDVP bond dimensions stay manageable; for the two-dimensional spin model, however, the truncation at $\chitdvp^{\max}=128$ is not negligible (Sec.~\ref{sec:heisenberg}). This provides a posteriori evidence for the numerical stability of the tensor-network benchmark in the instances considered here, consistent with the large initial overlap with the target low-energy eigenstate, rather than a general proof of stability across all problem instances.

\section{Auxiliary implementation-level resource estimates}
\label{sec:resources}
The benchmarks above use TDVP and do not depend on how $e^{-it\hat H\otimes\hat p}$ would be compiled. Here we give auxiliary implementation-level cost estimates for a possible Suzuki--Trotter realization of the protocol. Fault-tolerant cost bounds for the von Neumann measurement procedure with block-encoded Hamiltonians have been derived in Ref.~\cite{Skelton2026vonNeumannQPE}, and the MPO representation used in our simulations can itself serve as the interface to linear-combination-of-unitaries circuits for such block encodings~\cite{javanmard2026coefficientdecoupledmatrixproductoperators}; here we consider the simpler product-formula realization. Recent quantum-error-corrected demonstrations of molecular-energy estimation highlight the growing relevance of phase-estimation workflows in early fault-tolerant settings, while also illustrating the substantial resource demands of such implementations~\cite{Yamamoto2026QECMolecularEnergies}.

\subsection{Suzuki--Trotter decomposition}
\label{sec:trotter}
With the Hamiltonian in the form of Eq.~\eqref{eq:H_pauli} or Eq.~\eqref{eq:coupling}, the evolution operators are discretized with the first-order Suzuki--Trotter formula,
\begin{align}
e^{-i\hat Ht}&\approx\Big(\prod_{i}e^{-ic_i\hat h_i\delta t}\Big)^{n},\nonumber\\
e^{-it\hat H\otimes\hat p}&\approx\Big(\prod_{i}\prod_{j}e^{-ic_i\hat h_i\hat p_j\delta t}\Big)^{n},
\label{eq:trotter}
\end{align}
where $t=n\delta t$, $n$ is the number of Trotter steps, and $\delta t$ is the Trotter step. In a standard circuit implementation, each exponential is decomposed into single-qubit rotations and CNOT gates; Fig.~\ref{fig:gates} shows examples.

\subsection{Logical gate counts and compilation model}
\label{subsec:logical_gate_counts}
\begin{figure}[t]
\centering
\includegraphics[width=\textwidth]{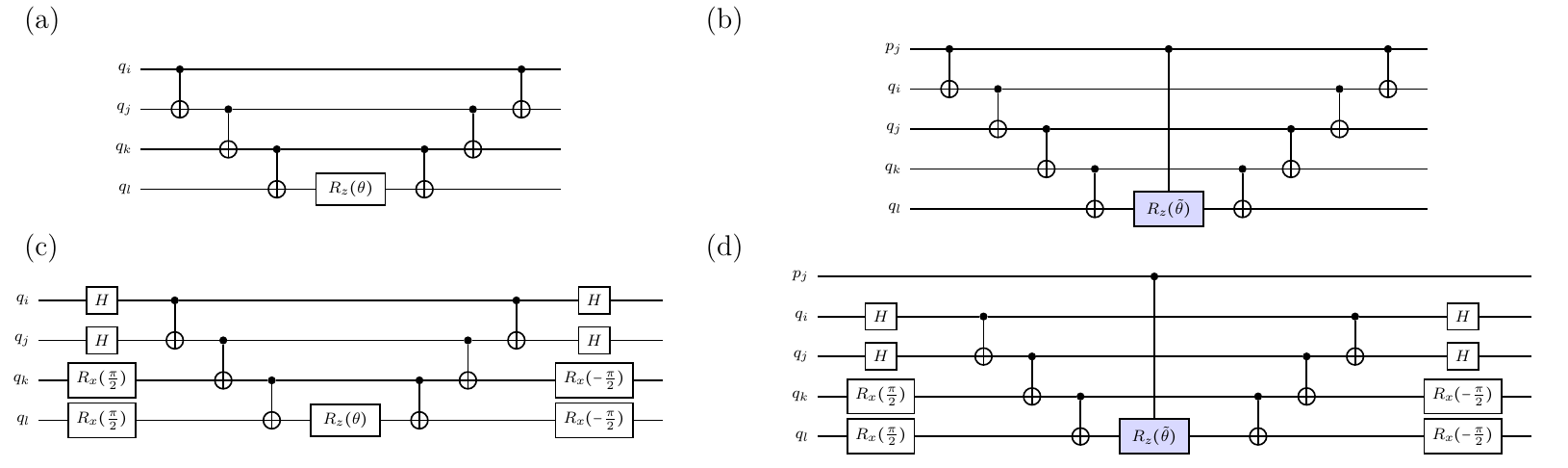}
\caption{Quantum circuits for the exponentials of Pauli strings used in the gate counts of Sec.~\ref{sec:resources}. (a)~$e^{-i\frac{\theta}{2}Z_iZ_jZ_kZ_l}$: CNOT ladder around a single $R_z(\theta)$. (b)~$e^{-i\frac{\theta}{2}Z_iZ_jZ_kZ_l\,p_j}$: the same term coupled to pointer qubit $j$. (c)~$e^{-i\frac{\theta}{2}X_iX_jY_kY_l}$: as (a) with single-qubit basis changes. (d)~$e^{-i\frac{\tilde\theta}{2}X_iX_jY_kY_l\,p_j}$ with $\tilde\theta=2^{-j-1}\theta$. The long-range controls are logical; on hardware with restricted connectivity they require SWAP routing (Sec.~\ref{subsec:connectivity}).}
\label{fig:gates}
\end{figure}

\paragraph*{Logical counting model.} We report logical gate counts in a standard gate library (CNOT and single-qubit rotations), using the circuit templates of Fig.~\ref{fig:gates}. In this model, a $k$-local Pauli-string exponential is implemented by basis changes and a CNOT ladder, requiring $2(k-1)$ CNOT gates and at most $2k+1$ single-qubit gates.

Many of these circuits have the conjugation pattern
\begin{equation}
U_c=VAV^\dagger ,
\label{eq:conjugation_pattern}
\end{equation}
for which it suffices to control the central operation $A$:
\begin{equation}
C[U_c]=C[VAV^\dagger]=(I\otimes V)\,C[A]\,(I\otimes V^\dagger),
\label{eq:controlled_conjugation}
\end{equation}
with the controlled operation
\begin{equation}
C[P]=\ket0\!\bra0\otimes I+\ket1\!\bra1\otimes P .
\label{eq:controlled_P}
\end{equation}
For circuit identities and compilation patterns for controlled operations we refer to Refs.~\cite{Cowtan_2020, Sivarajah_2021, peng_quantum_2022}.

\paragraph*{Controlled overhead from the pointer coupling.} In addition to the system evolution, the pointer coupling introduces controlled rotations: for each Trotter step, $r\times K$ controlled-$R_z(\theta)$ operations are required (pointer size $r$, number of Pauli terms $K$). By Eqs.~\eqref{eq:conjugation_pattern}--\eqref{eq:controlled_conjugation}, only the central $R_z(\theta)$ must be controlled, while the basis changes and CNOT ladders remain uncontrolled. Equivalently, since $\hat p_j=2^{-j}(1-\hat Z_j)/2$, each factor in Eq.~\eqref{eq:trotter} splits into an uncontrolled system rotation and a rotation of the Pauli string extended by $\hat Z_j$ on pointer qubit $j$ [Fig.~\ref{fig:gates}(b,d)]. Each controlled-$R_z$ decomposes into a constant number of native two-qubit gates, e.g.\ two CNOTs plus single-qubit rotations, so the pointer coupling adds $G_{\mathrm{ctrl}}\,rK$ two-qubit gates per step with $G_{\mathrm{ctrl}}=2$.

\subsection{Per-step costs and total cost up to time $t=n\delta t$}
\label{subsec:total_cost}
Table~\ref{tab:cnots} lists the logical CNOT counts for one Trotter step of the two molecular benchmarks, obtained by compiling the Pauli-term decomposition of the corresponding $\mathbb Z_2$-tapered 13-qubit Hamiltonians. These counts are instance specific: changing the active space changes the number of qubits and the number and structure of Pauli terms, and hence the counts. They are conservative logical estimates within the chosen Pauli-term and product-formula compilation model. Quantum-chemistry Hamiltonians are resource intensive for essentially all known digital quantum-simulation approaches, and substantial reductions may be possible by optimizing the active space, fermion-to-qubit mapping, symmetry tapering, Hamiltonian factorization, product formula, routing, and gate synthesis; a full optimization is outside the scope of this work.

\begin{table}[t]
\caption{Logical CNOT counts per first-order Trotter step for the molecular benchmarks. $G^{\mathrm{sim}}_{\mathrm{step}}$: trotterized evolution under $\hat H$; $G_{\mathrm{ctrl}}rK$: additional CNOTs from the pointer coupling ($G_{\mathrm{ctrl}}=2$, $r=4$). Not included: MPS state preparation, the pointer QFT, measurement repetitions, and hardware routing overhead.}
\label{tab:cnots}
\begin{ruledtabular}
\begin{tabular}{lcccc}
Molecule & Pauli terms $K$ & $G^{\mathrm{sim}}_{\mathrm{step}}$ & $G_{\mathrm{ctrl}}rK$ & Total per step \\
\hline
$\mathrm{H}_8$ (STO-3G, FCI) & 2912 & 37{,}500 & 23{,}296 & 60{,}796 \\
Pyridine (STO-3G, (8e,8o)) & 2764 & 30{,}444 & 22{,}112 & 52{,}556 \\
\end{tabular}
\end{ruledtabular}
\end{table}

The total two-qubit gate count of a single circuit execution up to $t=n\delta t$ is
\begin{equation}
G_{\mathrm{tot}}\approx n\big(G^{\mathrm{sim}}_{\mathrm{step}}+G_{\mathrm{ctrl}}\,rK\big)+G_{\mathrm{QFT}}(r)+G_{\mathrm{prep}}(\chi)+G_{\mathrm{route}},
\label{eq:total_cost}
\end{equation}
where $G_{\mathrm{QFT}}(r)$ is the cost of the inverse QFT on the pointer, $G_{\mathrm{prep}}(\chi)$ that of the MPS-loading circuit (Sec.~\ref{sec:mps_loading}), and $G_{\mathrm{route}}$ the routing overhead for restricted connectivity. An inverse QFT on $r$ qubits uses $\mathcal O(r^2)$ controlled-phase gates, about $r(r-1)/2$ plus $r$ Hadamards, i.e.\ 6 controlled phases for $r=4$. As an example, the $\mathrm{H}_8$ benchmark reaches chemical accuracy at $t\approx25\,\mathrm{Ha}^{-1}$ (Fig.~\ref{fig:convergence}); at $\delta t=0.05$, i.e.\ $n=500$, the evolution block alone amounts to about $500\times60{,}800\approx3\times10^{7}$ logical CNOTs per circuit execution in this model, and to about $500\times25{,}200\approx1.3\times10^{7}$ after the optimized compilation of Sec.~\ref{subsec:compilers}. For large $t$ the cumulative logical cost thus becomes very substantial, and these estimates should be viewed as logical scaling indicators rather than as evidence of practicality on present-day devices.

A further limitation of the logical product-formula model is that, at finite $\delta t$, the circuit implements evolution under an effective Hamiltonian $\widetilde H_{\delta t}$ rather than $\hat H$. Even without hardware noise, the energies inferred from the pointer can therefore differ from those of the target Hamiltonian by a digital simulation bias that depends on the product formula, the commutator structure, and the step size (Appendix~\ref{app:trotter_error_analysis}). The gate counts should be read as a logical compilation model, not as a certification of digital accuracy.

\paragraph*{Initial-state preparation cost.} The cost $G_{\mathrm{prep}}(\chi)$ of preparing the DMRG/MPS initial state depends on the loading circuit, the target fidelity, and the compilation strategy. It increases with the system size and with the bond dimensions $\{\chi_j\}$ of the MPS, and may become significant when high-fidelity loading of large-bond-dimension states is required. For sequential isometry-based loading, $G_{\mathrm{prep}}$ grows polynomially: linearly in the number of sites at fixed $\chi$, with a prefactor that grows with $\chi$ because the unitaries $U_i$ act on $1+\lceil\log_2\chi\rceil$ qubits. In the illustrative $\chi=2$ case of Fig.~\ref{fig:mps_loading}, the loading uses a single staircase of nearest-neighbor two-qubit gates. For the benchmark bond dimensions ($\chidmrg\le20$, i.e., unitaries on at most 6 qubits), the preparation cost is small compared with the evolution block, but we report it separately as a compilation-dependent term.

\subsection{Optimized compilation with current compilers}
\label{subsec:compilers}
The counts of Table~\ref{tab:cnots} assume one CNOT ladder per Pauli string and no optimization between strings. Current compilers reduce them substantially by synthesizing the whole sequence of Pauli rotations at once: consecutive rotations share basis changes and parts of their CNOT ladders. To quantify this for our systems, we compiled one first-order Trotter step of $\hat H\otimes\hat p$ ($r=4$) for the Heisenberg model and for the tapered $\mathrm{H}_8$ and pyridine Hamiltonians with three tool chains: Qiskit~\cite{JavadiAbhari2024Qiskit} at optimization level 3; Qiskit with the Rustiq synthesis of Pauli-rotation sequences~\cite{deBrugiere2024Rustiq}; and TKET~\cite{Sivarajah_2021} with greedy Pauli-gadget simplification followed by full peephole optimization. Each factor of Eq.~\eqref{eq:trotter} is written as the rotation of the system string $\hat h_i$ and the rotations of the strings $\hat h_i\hat Z_j$ ($j=1,\dots,r$), so a step contains $(r+1)K$ Pauli rotations. Each compiled circuit was verified against the exact product of Pauli rotations by state-vector simulation of a random input state.

\begin{table}[t]
\caption{Two-qubit (CNOT) gate counts for one first-order Trotter step of $\hat H\otimes\hat p$ with an $r=4$ pointer. Model: counting model of Sec.~\ref{subsec:logical_gate_counts}, $G^{\mathrm{sim}}_{\mathrm{step}}+G_{\mathrm{ctrl}}rK$ (Table~\ref{tab:cnots}). Qiskit~2.5 (optimization level 3), Qiskit with Rustiq synthesis, and TKET (pytket~2.18, greedy Pauli simplification and full peephole optimization) assume all-to-all connectivity. Counts marked $^{*}$ belong to circuits in which the compiler dropped very small rotation angles (infidelity $10^{-8}$--$10^{-6}$ for a random input); the value in parentheses is the count obtained when all angles are scaled to order one, i.e.\ the reduction due to the circuit structure alone. All other circuits reproduce the exact Trotter step to an infidelity below $10^{-11}$. The molecular Hamiltonians are the $\mathbb Z_2$-tapered 13-qubit ones.}
\label{tab:compiled}
\begin{ruledtabular}
\begin{tabular}{lccccc}
System (qubits incl.\ pointer, $K$) & Model & Qiskit O3 & Rustiq & TKET \\
\hline
Heisenberg $4\times3$ (16, 108) & 1{,}080 & 1{,}944 & 1{,}008 & 866 \\
$\mathrm{H}_8$ (17, 2912) & 60{,}796 & 90{,}063$^{*}$ & 25{,}413$^{*}$ (44{,}160) & 25{,}158 \\
Pyridine (17, 2764) & 52{,}556 & 16{,}184$^{*}$ & 5{,}838$^{*}$ (25{,}614) & 23{,}723$^{*}$ (24{,}083) \\
\end{tabular}
\end{ruledtabular}
\end{table}

Table~\ref{tab:compiled} summarizes the results. TKET gives the lowest structural counts, i.e.\ counts that do not rely on dropping small rotations: relative to the counting model, the cost per step is reduced by a factor of about $2.4$ for $\mathrm{H}_8$, $2.2$ for pyridine, and $1.25$ for the Heisenberg model, whose two-local terms leave little to share. The default Qiskit synthesis does not exploit this structure and can exceed the counting model. The lowest numbers in the table (Rustiq for pyridine) are partly obtained by removing rotations with very small angles, which are numerous in molecular Hamiltonians at $\delta t=0.05$; such reductions are an approximation of the Trotter step, and we therefore list them separately from the structural ones. Screening of small Pauli coefficients is a legitimate further optimization, but its effect on the energy must then be accounted for in the error budget.

The compiled counts assume all-to-all connectivity; on hardware with restricted connectivity, the routing overhead $G_{\mathrm{route}}$ of Eq.~\eqref{eq:total_cost} must be added (Sec.~\ref{subsec:connectivity}).

\subsection{Connectivity and hardware realism}
\label{subsec:connectivity}
The circuits of Fig.~\ref{fig:gates} depict logical control structures. On hardware with restricted connectivity, e.g.\ nearest-neighbor layouts, long-range two-qubit gates require SWAP routing, which increases both gate count and depth. On a one-dimensional nearest-neighbor chain, with one SWAP equal to 3 CNOTs, a two-qubit gate between qubits at graph distance $d$ can be implemented by moving one qubit next to the other and back, at a cost of about $2(d-1)$ SWAPs, i.e.\ $6(d-1)$ additional CNOTs, up to layout- and compiler-dependent constants. To avoid over-interpreting logical counts, we keep this hardware-dependent overhead separate as $G_{\mathrm{route}}$ in Eq.~\eqref{eq:total_cost}.

\paragraph*{SWAP-free routing with Parity Twine.} Routing overhead can be avoided for important subroutines. Parity Twine networks implement circuits built from two-qubit interactions on linear nearest-neighbor hardware without SWAP gates and without additional qubits~\cite{Klaver2026SWAPless}. For the QFT on $n$ qubits they require $n^2-1$ CNOT gates and a circuit depth of $5n-3$, improving on previous linear-nearest-neighbor implementations of the QFT~\cite{Klaver2026SWAPless}; adaptations to neutral-atom layouts and native gate sets further reduce gate count, atom transport, and depth~\cite{Kazemi2026ParityTwine}. For the pointer register this gives $G_{\mathrm{QFT}}(r)=r^2-1=15$ CNOTs at depth $5r-3=17$ for $r=4$, negligible compared with the evolution block, and keeps the readout stage free of routing overhead as $r$ is increased for higher energy resolution. The same construction applies to Hamiltonians with two-body terms, such as the Heisenberg model; extending SWAP-free schemes to the high-weight Pauli strings of molecular Hamiltonians is beyond the scope of this work.

\paragraph*{Noise sensitivity.} We do not perform device-level noise simulations, and the discussion in this section is a logical compilation analysis rather than a performance prediction for present-day devices. The dominant hardware sensitivities are expected to come from the two-qubit-gate count of the controlled trotterized evolutions, the cumulative depth over many time steps, and the overhead of state preparation, routing, and readout; these effects reduce the visibility of the peaks in the pointer distribution and thereby the precision of the fitted energy.

\section{Discussion}
\label{sec:discussion}
\paragraph*{What the benchmarks show.} For all four Hamiltonians, the pointer readout of the von Neumann measurement protocol recovers the ground-state energy from DMRG initial states of small bond dimension ($\chidmrg=2$--$20$) whose own energies are far from converged, and for the molecules it does so within chemical accuracy at evolution times far shorter than the conservative single-shot bound. These small bond dimensions reflect the size of the instances: the bond dimension needed for a given overlap with the ground state grows with system size and with the strength of correlations, so larger molecules or lattices will require correspondingly larger $\chidmrg$, and the TDVP bond dimension needed for the coupled evolution grows as well. The time-resolved pointer distribution has a transparent structure---periodic peak trains with revival time $2^r\cdot2\pi/|E_0|$---so that every row provides an independent estimate and deviating rows are easy to identify. The system--pointer entanglement provides a simulation-internal measure of the initial-state quality.

\paragraph*{Choice of the initial state in practice.} The bond dimension of the DMRG input is a tunable resource. Increasing $\chidmrg$ raises the overlap with the ground state, so that the pointer distribution is less contaminated by excited-state components: the peaks are cleaner, fewer repetitions are needed, and the fitted energy is less affected by other eigencomponents. This can reduce the evolution time, and hence the number of Trotter steps, required for a given accuracy. The price is a deeper loading circuit, whose unitaries act on $1+\lceil\log_2\chi\rceil$ qubits~\cite{Ran20, rudolph2022decomposition}, and a larger classical cost of the DMRG calculation. In practice, $\chidmrg$ can therefore be chosen large enough to capture the dominant correlations of the target state while keeping the loading circuit shallow compared with the controlled time evolution. Other state-preparation methods can be used in the same way, for example variational circuits optimized with the variational quantum eigensolver~\cite{peruzzoVariationalEigenvalueSolver2014}, possibly pre-optimized with tensor networks~\cite{khanPreoptimizingVariationalQuantum2023}, adiabatic state preparation~\cite{farhiQuantumAdiabaticEvolution2001}, or compressed imaginary-time circuits~\cite{Lin21}. The simulation framework applies to any input state that can be represented as an MPS.

\paragraph*{Limitations.} (i) The instances are small enough that exact reference energies are available, which is necessary for a benchmark but does not probe the regime where DMRG itself fails. (ii) Trotter error and hardware noise are not simulated (Sec.~\ref{sec:tn_scope}). (iii) We make no claim of an advantage over textbook QPE, over modern fault-tolerant phase-estimation methods, or over VQE; the protocol trades coherent depth, number of repetitions, and ancilla count differently, and the preferable trade-off depends on the hardware and accuracy target. (iv) The quality of the initial states is characterized here by their energy error and by the system--pointer entanglement they generate. (v) The TDVP simulations start from an unsaturated bond dimension, which introduces a projection error of the order of $10^{-2}$ in the pointer distribution (Sec.~\ref{sec:tn_scope}); for the Heisenberg benchmark it accounts for about $4\times10^{-3}\,J$ of the deviation from $E_0$, and the truncation at $\chitdvp^{\max}=128$ contributes further (Sec.~\ref{sec:heisenberg}). The reported deviations therefore include simulation error and do not isolate the error of the protocol. (vi) The energy fit uses a prior window below $E_{\mathrm{init}}$ that must contain the target energy; for the benchmarks, the recovered energies are unchanged when the window is widened from one to three units, and no row estimate lies on the window edge.

\subsection{Relation to hybrid tensor-network quantum simulation}
\label{sec:relation_hybrid_tn}
Several works have proposed \emph{hybrid} tensor-network quantum simulation, in which tensor networks reduce the hardware qubit requirement by combining small quantum-prepared or quantum-measured building blocks with classical tensors contracted on a classical computer~\cite{Yuan2021PRLHybridTN, Yuan2020arxivHybridTN}. Our use of tensor networks is different. We assume that the full system register fits on the device, together with a small pointer register, and use a pre-trained DMRG/MPS state as a high-overlap initializer; the energy is then extracted from the pointer distribution of the von Neumann measurement, e.g.\ by fitting. In addition, tensor networks serve here as the classical simulator used to benchmark the protocol. As a simulator, the framework evolves the Hamiltonian $\hat H\otimes\hat p$ rather than a gate sequence, which complements TDVP-based simulation of compiled circuits~\cite{sander2025quantumcircuitsimulationlocal}.

\section{Conclusion}
\label{sec:conclusion}
We have presented a tensor-network framework for benchmarking the von Neumann measurement formulation of quantum phase estimation, introduced by Childs \emph{et al.}~\cite{Qsearchmeasurement2002} and analyzed for many-body energy measurements by Novo \emph{et al.}~\cite{novo_quantum_2021}, with DMRG-pretrained MPS as initial states. Representing the Hamiltonian and states as MPOs and MPSs, the coupling to the pointer as an MPO without bond-dimension overhead, and the coupled evolution with TDVP gives direct access to the pointer distribution, the recovered energy, and the system--pointer entanglement.

We applied the framework to a strongly correlated, frustrated spin system and to electronic-structure problems of octahydrogen ($\mathrm{H}_8$), pyridine ($\mathrm{C_5H_5N}$), and $\mathrm{BH}_3$. Within the tensor-network benchmark model, the pointer readout recovers accurate ground-state energies from poor DMRG inputs, reaching chemical accuracy for the molecules, and the system--pointer entanglement shows how close the input is to an eigenstate. The corresponding digital circuit realization was analyzed at the level of logical compilation and resource estimation; a full end-to-end assessment of product-formula error and hardware noise remains for future work.

The approach can be useful in quantum chemistry and materials science, where accurate simulation of quantum systems is essential. A natural experimental direction is suggested by recent demonstrations of phase estimation for molecular energies on trapped-ion hardware with error-detected~\cite{Yamamoto2024BayesianQPE} and error-corrected~\cite{Yamamoto2026QECMolecularEnergies} logical qubits: running the von Neumann measurement protocol with DMRG-prepared inputs on such logical-qubit hardware, with the tensor-network framework used to design and benchmark the circuits beforehand. Further directions include variational implementations of the protocol~\cite{filip_variational_2024, liu2023learning, Cirstoiu2020}, the inclusion of product-formula error and noise in the tensor-network simulation, and instances in which DMRG alone is not converged. In particular, the framework can be extended from the ideal evolution under $\hat H\otimes\hat p$ to the simulation of the compiled quantum circuits themselves, i.e.\ the Trotterized controlled evolutions and the pointer QFT of Sec.~\ref{sec:resources}, using a local TDVP for quantum circuits as developed in Ref.~\cite{sander2025quantumcircuitsimulationlocal}; this would give direct access to the product-formula error and to the effect of finite circuit depth on the pointer readout. Appendix~\ref{app:excited} outlines a possible extension of the framework to excited states; the validated benchmarks and claims of the present work are restricted to ground-state energy estimation.

\begin{acknowledgments}
The author thanks Tobias J.~Osborne for insightful discussions and for suggesting the core direction of this work. We acknowledge funding by the Ministry of Science and Culture of Lower Saxony through \textit{Quantum Valley Lower Saxony Q1} (QVLS-Q1).
\end{acknowledgments}

\paragraph*{Code and data availability.} The code used in this work, the simulation data underlying the figures, and scripts that reproduce the figures and the validation of Sec.~\ref{sec:tn_scope} are publicly available at \url{https://github.com/younesjavanmard/vn-qpe-tn-benchmark}.

\paragraph*{Use of AI tools.} Claude (Anthropic) and ChatGPT (OpenAI) were used for code assistance and refactoring, data analysis, and language editing. All content was verified by the author.

\appendix

\section{Jordan--Wigner transformation}
\label{app:jw}
The Jordan--Wigner transformation maps $M$ fermionic modes onto $M$ ordered qubits. The value of qubit $j$ encodes the occupation of mode $j$, and the parity of the occupations of the modes preceding $j$ is stored in a string of $\hat Z$ operators on the corresponding qubits:
\begin{equation}
f_j=\Big(\prod_{i=1}^{j-1}\hat Z_i\Big)\sigma^+_j,\qquad f_j^\dagger=\Big(\prod_{i=1}^{j-1}\hat Z_i\Big)\sigma^-_j,
\end{equation}
where $\sigma^\pm_j=(\hat X_j\pm i\hat Y_j)/2$. The fermionic operators satisfy the canonical anticommutation relations
\begin{equation}
\{f_i,f_j\}=0,\qquad\{f_i^\dagger,f_j^\dagger\}=0,\qquad\{f_i,f_j^\dagger\}=\delta_{ij}.
\end{equation}
The $\hat Z$ strings make the molecular Hamiltonians long-ranged, which is why their MPOs are constructed with the graph-based method of Sec.~\ref{sec:tn_mpo}~\cite{javanmard2026coefficientdecoupledmatrixproductoperators}.

\section{Graph-based MPO construction for molecular Hamiltonians}
\label{app:mpo_graph}
\begin{figure}[h]
\centering
\includegraphics[width=\textwidth]{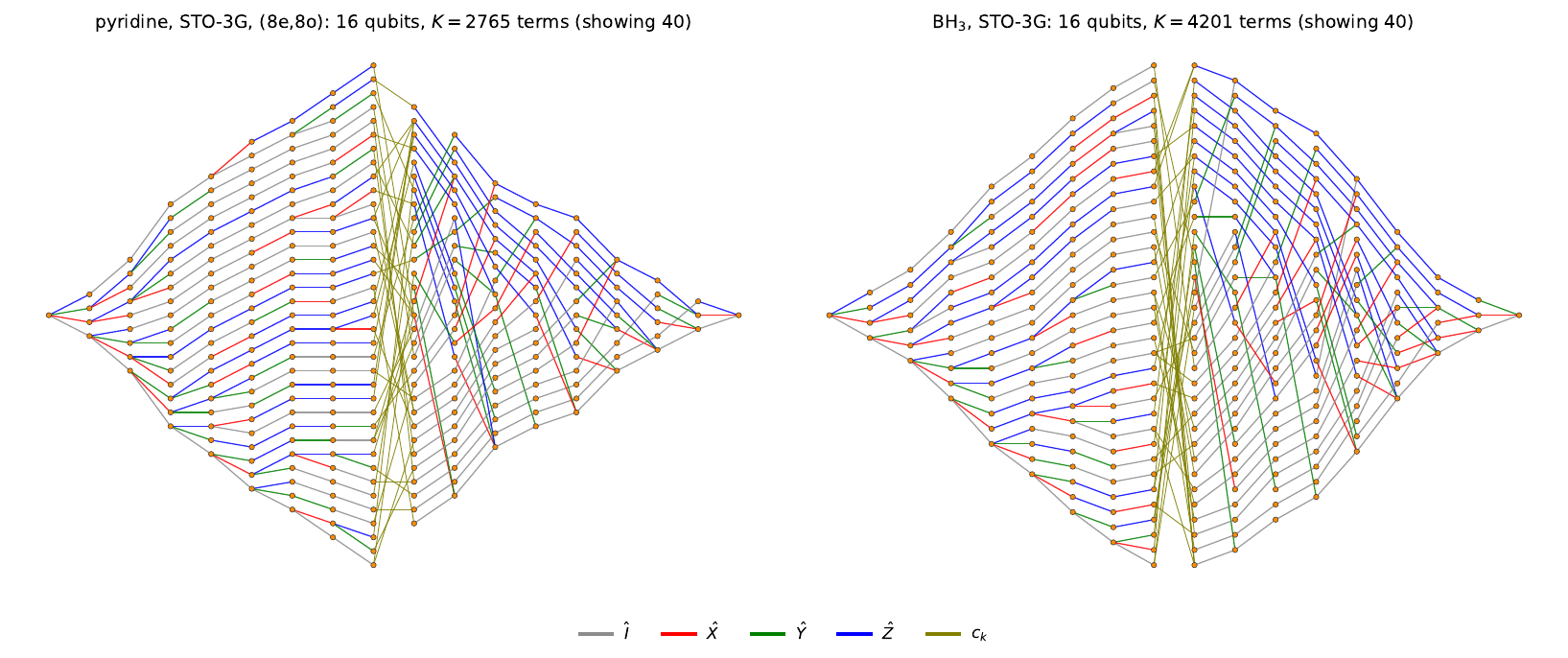}
\caption{Graph-based MPO construction for two molecular Hamiltonians of this work: (left) pyridine, STO-3G, (8e,8o) active space; (right) $\mathrm{BH}_3$, STO-3G. Both are Jordan--Wigner Hamiltonians on 16 qubits. For visibility, each panel shows a random subset of 40 of the $K$ Pauli strings. Vertices are distinct Pauli-string fragments, arranged in layers by the site at which the fragment ends; edges are labeled by the single-site operator that extends a fragment by one site (colors in the legend); olive edges carry the coefficients $c_k$ at the central cut. Conclusion: fragments shared by many strings are merged, so the MPO bond dimension of the full Hamiltonians ($D_H=138$ and $133$ after compression) is far below the number of terms ($K=2765$ and $5437$).}
\label{fig:mpo_graph}
\end{figure}

Figure~\ref{fig:mpo_graph} illustrates the graph-based construction of Sec.~\ref{sec:tn_mpo}~\cite{javanmard2026coefficientdecoupledmatrixproductoperators} for two molecules of this work, pyridine in the (8e,8o) active space (STO-3G) and $\mathrm{BH}_3$ (STO-3G), both mapped to 16 qubits with the Jordan--Wigner transformation. Each Pauli string is split at the central cut into a left fragment (a prefix) and a right fragment (a suffix). Identical fragments shared by several strings are merged into a single vertex, and the coefficients $c_k$ are attached to the edges that connect the left and right graphs at the cut. Reading the graph layer by layer gives the MPO tensors, and the number of vertices in a layer is the MPO bond dimension on that bond before compression.

For the full Hamiltonians, the pyridine active space has $K=2765$ Pauli strings and $\mathrm{BH}_3$ has $K=4201$. The graph gives maximal MPO bond dimensions of $437$ and $499$, respectively, which the numerically lossless compression of Sec.~\ref{sec:tn_mpo} reduces to $D_H=138$ and $D_H=154$, i.e., more than an order of magnitude below the number of terms. For $\mathrm{BH}_3$, the geometry of Table~\ref{tab:geometries} admits several Hartree--Fock solutions, and $K$ and $D_H$ depend on the converged orbitals (we find $K$ between about 3100 and 5500 in different runs); the FCI energy of Appendix~\ref{app:BH3} does not depend on this choice. The coupling to the pointer appends the bond-dimension-2 chain of Fig.~\ref{fig:mpo_coupled} to the right end of this graph and does not change $D_H$.

\section{Molecular geometries}
\label{app:geometry}
Table~\ref{tab:geometries} lists the Cartesian coordinates of the molecules used in this work.

\begin{table}[h]
\caption{Geometries of $\mathrm{H}_8$ (left), pyridine (center), and $\mathrm{BH}_3$ (right).}
\label{tab:geometries}
\begin{ruledtabular}
\begin{tabular}{lrrr}
\multicolumn{4}{c}{$\mathrm{H}_8$}\\
\hline
 & $x$ & $y$ & $z$ \\
H & 1.6180339887 & 0.0000000000 & 0 \\
H & 1.3090169944 & 0.9510565163 & 0 \\
H & 0.5000000000 & 1.5388417686 & 0 \\
H & $-0.5000000000$ & 1.5388417686 & 0 \\
H & $-1.3090169944$ & 0.9510565163 & 0 \\
H & $-1.6180339887$ & 0.0000000000 & 0 \\
H & $-1.3090169944$ & $-0.9510565163$ & 0 \\
H & $-0.5000000000$ & $-1.5388417686$ & 0 \\
\end{tabular}
\hspace{0.03\textwidth}
\begin{tabular}{lrrr}
\multicolumn{4}{c}{Pyridine}\\
\hline
 & $x$ & $y$ & $z$ \\
C & 1.3603 & 0.0256 & 0 \\
C & 0.6971 & $-1.2020$ & 0 \\
C & $-0.6944$ & $-1.2184$ & 0 \\
C & $-1.3895$ & $-0.0129$ & 0 \\
C & $-0.6712$ & 1.1834 & 0 \\
N & 0.6816 & 1.1960 & 0 \\
H & 2.4530 & 0.1083 & 0 \\
H & 1.2665 & $-2.1365$ & 0 \\
H & $-1.2365$ & $-2.1696$ & 0 \\
H & $-2.4837$ & 0.0011 & 0 \\
H & $-1.1569$ & 2.1657 & 0 \\
\end{tabular}
\hspace{0.03\textwidth}
\begin{tabular}{lrrr}
\multicolumn{4}{c}{$\mathrm{BH}_3$}\\
\hline
 & $x$ & $y$ & $z$ \\
B & 0 & 0 & 0 \\
H & 0 & 1 & 1 \\
H & 1 & 0 & 1 \\
H & 1 & 1 & 0 \\
\end{tabular}
\end{ruledtabular}
\end{table}

\section{Additional benchmark: $\mathrm{BH}_3$}
\label{app:BH3}
\begin{figure}[h]
\centering
\includegraphics[width=0.45\textwidth]{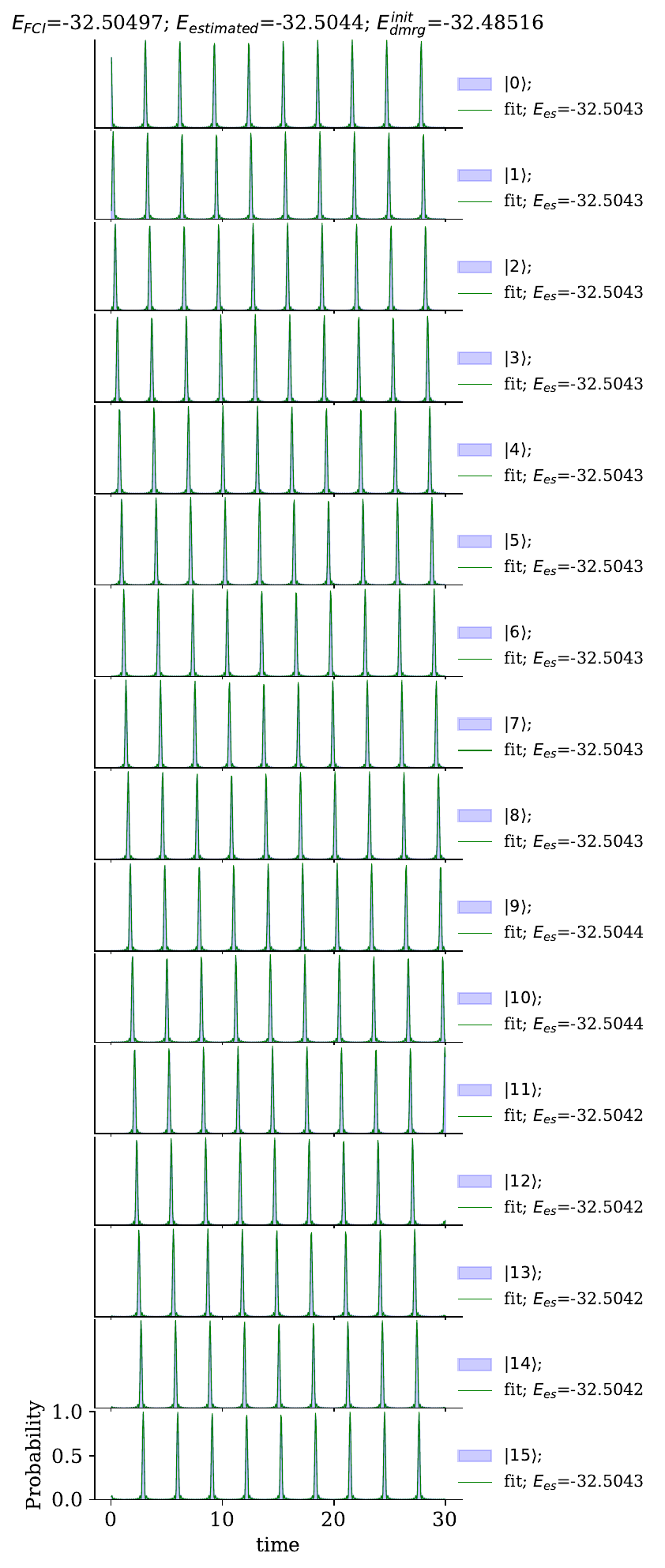}
\caption{$\mathrm{BH}_3$, STO-3G, FCI ($N=16$, $r=4$); electronic energies without the nuclear repulsion. Each panel shows $P(x;t)$ for pointer outcome $\ket{x}$ ($x=0,\dots,15$, top to bottom) as a function of the evolution time $t$ ($\mathrm{Ha}^{-1}$). Blue: TDVP data; green line: fit, with the resulting estimate $E_{\mathrm{es}}$ in the legend. Initial state: DMRG MPS with $\chidmrg=6$, $E^{\mathrm{init}}_{\mathrm{dmrg}}=-32.48516$\,Ha. TDVP: $\delta t=0.05\,\mathrm{Ha}^{-1}$, two-site TDVP until the bond dimension reaches 250, then one-site TDVP, $\chitdvp^{\max}=300$. Header: $E_{\mathrm{FCI}}$, recovered energy $E_{\mathrm{estimated}}$, and DMRG initial-state energy. The header shows the lowest row estimate. Conclusion: the recovered energy (median of the rows, $-32.5043$\,Ha) lies $0.67$\,mHa from $E_{\mathrm{FCI}}$.}
\label{fig:BH3}
\end{figure}

As a further molecular example, we consider $\mathrm{BH}_3$ in the STO-3G basis with the geometry of Table~\ref{tab:geometries}. It has 8 spatial orbitals, corresponding to 16 qubits, and is treated at the FCI level. Unlike the other molecules, the energies of this example are electronic energies, without the nuclear repulsion $E_{\mathrm{nuc}}=6.7353$\,Ha (total FCI energy $-25.7696$\,Ha); $E_{\mathrm{FCI}}=-32.50497$\,Ha. The DMRG initial state with $\chidmrg=6$ has $E_{\mathrm{init}}=-32.48516$\,Ha, i.e., $19.8$\,mHa above $E_{\mathrm{FCI}}$. Figure~\ref{fig:BH3} shows the time-resolved pointer distribution up to $t=30\,\mathrm{Ha}^{-1}$ (revival time $T_{\mathrm{rev}}\approx3.1\,\mathrm{Ha}^{-1}$). All sixteen rows give $E_{\mathrm{es}}$ between $-32.5042$ and $-32.5044$\,Ha, and the recovered energy $E_{\mathrm{estimated}}=-32.5043$\,Ha (median of the rows) lies $0.67$\,mHa from $E_{\mathrm{FCI}}$, consistent with the benchmarks of Sec.~\ref{sec:molecules}.

\section{TDVP time evolution}
\label{app:tdvp}
This appendix documents the tensor-network time evolution used for the benchmark figures; it should be distinguished from the Suzuki--Trotter product-formula model used for the resource estimates. We evolve MPS with the TDVP~\cite{haegeman_time-dependent_2011, haegeman_unifying_2016}, in its two-site form, which allows the bond dimensions to grow up to $\chitdvp^{\max}$; we first describe one-site TDVP, which keeps the bond dimensions fixed and underlies the two-site scheme. Throughout, $\delta t=0.05$, $\chitdvp^{\max}=128$ for the spin model and $100$ for the molecules, and the Hamiltonian is the coupled operator $\hat H\otimes\hat p$ on $L=N+r$ sites, denoted $\hat{\mathcal H}$ below.

\subsection{Projected Schr\"odinger equation and one-site TDVP}
Let $\ket{\Psi(t)}$ be an MPS with fixed bond dimensions. The Dirac--Frenkel--McLachlan variational principle chooses the trajectory within the MPS manifold that minimizes
\begin{equation}
\big\|\,i\,\partial_t\ket{\Psi(t)}-\hat{\mathcal H}\ket{\Psi(t)}\big\|^2 ,
\end{equation}
which is equivalent to the projected time-dependent Schr\"odinger equation
\begin{equation}
i\,\frac{d}{dt}\ket{\Psi(t)}=\mathcal P_{T,\ket{\Psi(t)}}\,\hat{\mathcal H}\,\ket{\Psi(t)},
\label{eq:projected-TDSE}
\end{equation}
where $\mathcal P_{T,\ket{\Psi(t)}}$ is the orthogonal projector onto the tangent space of the MPS manifold at $\ket{\Psi(t)}$. In mixed canonical form, the tangent-space projector decomposes into local projectors on sites and bonds,
\begin{equation}
\mathcal P_{T,\ket\Phi}=\sum_{\ell=1}^{L}\mathcal P^{\mathrm{site}}_\ell-\sum_{\ell=1}^{L-1}\mathcal P^{\mathrm{bond}}_\ell ,
\label{eq:proj-split}
\end{equation}
with
\begin{align}
\mathcal P^{\mathrm{site}}_\ell&=\ket{\Phi^L_{\ell-1}}\!\bra{\Phi^L_{\ell-1}}\otimes\mathbb I_\ell\otimes\ket{\Phi^R_{\ell+1}}\!\bra{\Phi^R_{\ell+1}},\\
\mathcal P^{\mathrm{bond}}_\ell&=\ket{\Phi^L_\ell}\!\bra{\Phi^L_\ell}\otimes\ket{\Phi^R_{\ell+1}}\!\bra{\Phi^R_{\ell+1}}.
\end{align}
Here $\ket{\Phi^L_\ell}$ and $\ket{\Phi^R_\ell}$ are the orthonormal left and right block states and $\mathbb I_\ell$ is the identity on the physical space of site $\ell$.

Fixing the orthogonality center at site $\ell$ and inserting Eq.~\eqref{eq:proj-split} into Eq.~\eqref{eq:projected-TDSE} gives local equations of motion for the center-site tensor $M_\ell$ and the adjacent bond tensor $C_\ell$,
\begin{align}
\frac{d}{dt}M_\ell(t)&=-i\,H^{\mathrm{eff}}_\ell\,M_\ell(t), && \ell=1,\dots,L,\label{eq:1tdvp-forward}\\
\frac{d}{dt}C_\ell(t)&=+i\,\widetilde H^{\mathrm{eff}}_\ell\,C_\ell(t), && \ell=1,\dots,L-1.\label{eq:1tdvp-backward}
\end{align}
The effective Hamiltonians $H^{\mathrm{eff}}_\ell$ and $\widetilde H^{\mathrm{eff}}_\ell$ are obtained by contracting the MPO of $\hat{\mathcal H}$ with all MPS tensors except the local object being updated. For a time step $\delta t$, the local propagators are applied with a short Krylov/Lanczos approximation,
\begin{align}
M_\ell(t+\delta t)&\approx\exp\!\big(-iH^{\mathrm{eff}}_\ell\delta t\big)\,M_\ell(t),\\
C_\ell(t+\delta t)&\approx\exp\!\big(+i\widetilde H^{\mathrm{eff}}_\ell\delta t\big)\,C_\ell(t).
\end{align}
After updating $M_\ell$, a QR decomposition $M_\ell=\widetilde M_\ell C_\ell$ moves the orthogonality center to site $\ell+1$, and the sweep continues. The backward bond updates restore the mixed canonical form, so that one left-to-right and right-to-left sweep approximates $\ket{\Psi(t+\delta t)}\approx\exp(-i\mathcal P_{T,\ket{\Psi(t)}}\hat{\mathcal H}\delta t)\ket{\Psi(t)}$.

\subsection{Two-site TDVP}
To allow entanglement growth, two-site TDVP updates a neighboring pair of sites $(\ell,\ell+1)$ by evolving the merged two-site tensor,
\begin{equation}
M_{\ell,\ell+1}(t+\delta t)\approx\exp\!\big(-iH^{\mathrm{eff}}_{\ell,\ell+1}\delta t\big)\,M_{\ell,\ell+1}(t),
\label{eq:2tdvp-forward}
\end{equation}
where $H^{\mathrm{eff}}_{\ell,\ell+1}$ is obtained by contracting the MPO with all tensors except the two-site block. The updated tensor is split by a singular value decomposition, $M_{\ell,\ell+1}=U_\ell S_\ell V^\dagger_{\ell+1}$, and the singular values are truncated according to a tolerance and the maximal bond dimension $\chitdvp^{\max}$; the one-site tensors are then $M_\ell=U_\ell$ and $M_{\ell+1}=S_\ell V^\dagger_{\ell+1}$, up to the usual gauge choices.

A common alternative is a dynamic scheme that switches to one-site TDVP once $\chi_\ell$ has saturated, which avoids truncation at the price of a fixed variational manifold. All benchmarks of the main text use two-site TDVP throughout, truncated at $\chitdvp^{\max}$; the $\mathrm{BH}_3$ example of Appendix~\ref{app:BH3} switches to one-site TDVP once the bond dimension reaches 250; for a Heisenberg run with these settings (Appendix~\ref{app:sampling}), the accumulated discarded weight reaches $8\times10^{-3}$ and $\la\hat H\otimes\hat p\ra$ drifts by $6\times10^{-2}\,J$ up to $t=10/J$. For the benchmarks considered here, the pre-trained DMRG initial state is already close to the ground state, so the entanglement growth over the relevant evolution window is modest and the required bond dimensions remain manageable (Sec.~\ref{sec:entanglement}).

\section{Idealized error analysis for the von Neumann pointer readout}
\label{app:vN_error_analysis}
This appendix summarizes an idealized (algorithmic) precision analysis of the pointer readout, following Ref.~\cite{novo_quantum_2021}. We consider only effects intrinsic to a finite pointer register and a finite evolution time. TDVP truncation error, Suzuki--Trotter error (Appendix~\ref{app:trotter_error_analysis}), state-preparation error, and hardware noise are not included, so the analysis is a readout-resolution analysis rather than an end-to-end accuracy guarantee for a compiled circuit.

\subsection{Setup and estimator}
The pointer register has $r$ qubits, so the outcome is $x\in\{0,1,\dots,2^r-1\}$. For an eigenstate with energy $E$, the inverse QFT produces a distribution $P(x)$ sharply peaked around
\begin{equation}
x^\star\approx-\frac{E\,t}{2\pi}\pmod{2^r}.
\end{equation}
Given a measured outcome $\hat x$, we use the standard estimator
\begin{equation}
\hat E=-\frac{2\pi\,\hat x}{t}.
\label{eq:vN_E_hat_def}
\end{equation}

\paragraph*{Aliasing / energy-range condition.} Because the readout is defined modulo $2^r$, wrap-around is avoided if the relevant energy range fits into the phase window; a sufficient condition is
\begin{equation}
\frac{(E_{\max}-E_{\min})\,t}{2\pi}<2^r,
\label{eq:vN_no_aliasing}
\end{equation}
where $[E_{\min},E_{\max}]$ is a known spectral interval of interest or an upper bound. If needed, the Hamiltonian can be shifted by a known constant to reduce the required range.

\subsection{Acceptance window and failure probability}
We introduce an integer acceptance window of half-width $w>1$ around the peak,
\begin{equation}
|\hat x-x^\star|\le w .
\label{eq:vN_success_window}
\end{equation}
A standard phase-estimation tail bound~\cite{Nielsen_Chuang_2010} implies
\begin{equation}
\Pr(\text{success})\ge1-\frac{1}{2(w-1)}\equiv1-\delta,
\label{eq:vN_success_prob}
\end{equation}
so that $\delta\in(0,1)$ upper-bounds the single-shot failure probability. Solving for $w$ gives the convenient choice
\begin{equation}
w_\delta=\Big\lceil\frac{1}{2\delta}\Big\rceil+1 .
\label{eq:vN_w_of_delta}
\end{equation}

\subsection{Energy-estimation error bound}
If Eq.~\eqref{eq:vN_success_window} holds, then
\begin{equation}
|E-\hat E|=\Big|E+\frac{2\pi\hat x}{t}\Big|\le\frac{2\pi w}{t}.
\label{eq:vN_energy_error}
\end{equation}
To achieve precision $\epsilon>0$ with success probability at least $1-\delta$, it therefore suffices to choose $w=w_\delta$ and
\begin{equation}
t\ge\frac{2\pi w_\delta}{\epsilon}.
\label{eq:vN_t_for_epsilon}
\end{equation}

\subsection{Resolving neighboring eigenvalues}
Let $\Delta_k=\min_{j\ne k}|E_k-E_j|$ be the nearest-neighbor gap of level $k$. A sufficient condition to avoid confusing neighboring levels is
\begin{equation}
\frac{2\pi w_\delta}{t}<\Delta_k\qquad\Longleftrightarrow\qquad t>\frac{2\pi w_\delta}{\Delta_k}.
\label{eq:vN_t_for_gap}
\end{equation}
In practice, $t$ is chosen large enough to satisfy both Eq.~\eqref{eq:vN_t_for_epsilon} (precision) and Eq.~\eqref{eq:vN_t_for_gap} (resolvability), while respecting the no-aliasing condition, Eq.~\eqref{eq:vN_no_aliasing}.

\subsection{Boosting confidence by repetition}
Equation~\eqref{eq:vN_w_of_delta} shows that driving the single-shot failure probability $\delta$ to very small values is conservative, since it increases $w_\delta$ and hence the required $t$ at fixed $\epsilon$. Alternatively, one can keep a moderate $\delta$ and repeat the protocol $m$ times, combining the outcomes, e.g.\ by a median or majority rule; under mild independence assumptions the overall failure probability then decreases exponentially in $m$ without requiring $t$ to scale as $1/\delta$.

\subsection{Worked example: chemical accuracy}
\label{app:vN_chem}
For chemical accuracy,
\begin{equation}
\epsilon_{\mathrm{chem}}\equiv1\ \mathrm{kcal/mol}\approx1.5936\times10^{-3}\ \mathrm{Ha},
\end{equation}
and a moderate single-shot failure probability $\delta=0.1$, Eq.~\eqref{eq:vN_w_of_delta} gives $w_\delta=\lceil1/0.2\rceil+1=6$, and Eq.~\eqref{eq:vN_t_for_epsilon} gives the conservative requirement
\begin{equation}
t\ge\frac{2\pi w_\delta}{\epsilon_{\mathrm{chem}}}\approx2.37\times10^4\ \mathrm{Ha}^{-1}.
\end{equation}
This bound is worst case; in the benchmarks of Sec.~\ref{sec:molecules} the fitted energies reach chemical accuracy at much smaller $t$, consistent with the observed sharpening of the pointer distribution.

\section{Finite-shot pointer readout from perfect sampling of the MPS}
\label{app:sampling}
The benchmarks of Sec.~\ref{sec:benchmarks} use the exact pointer distribution $P(x;t)$ of Eq.~\eqref{eq:readout_tn}. In an experiment, $P(x;t)$ is only known through a finite number $N_s$ of measurement outcomes per time step. Here we quantify how this shot noise affects the energy estimate, for the Heisenberg benchmark of Fig.~\ref{fig:heisenberg}.

\paragraph*{Perfect sampling.} Measurement outcomes can be drawn directly from an MPS by perfect sampling~\cite{Ferris2012PerfectSampling}: the sites are sampled one after another from their conditional probabilities given the outcomes already drawn, which are obtained by contracting the MPS with precomputed environments. Each sample costs $\mathcal O(L\chi^3)$ operations, is exact, and is independent of all previous samples, so no Markov chain, equilibration, or autocorrelation is involved. For the pointer readout, the inverse QFT is first applied to the $r$ pointer sites, and the system is traced out exactly through its left environment (Sec.~\ref{sec:tn_readout}); only the $r$ pointer sites are then sampled. For $r=4$ the resulting outcomes have exactly the statistics of independent draws from the 16-outcome distribution $P(x;t)$, which we use below; the sampler becomes necessary for large pointers, where $P(x;t)$ cannot be tabulated. Our implementation reproduces the exact distribution of random MPS (chi-squared per degree of freedom $0.96$--$1.15$ for $4\times10^5$ samples). We also tested it on a time-evolved state of the same benchmark, obtained with the same TDVP settings from a more converged $\chidmrg=20$ input ($E_{\mathrm{init}}=-26.988\,J$): drawing $10^5$ samples from the MPS at each of the 200 time steps gives histograms consistent with the exact pointer distribution of that MPS (mean chi-squared per degree of freedom $0.96$, maximal total-variation distance $5.8\times10^{-3}$), and the distribution of the per-step $p$ values is consistent with uniform ($p=0.20$, Kolmogorov--Smirnov).

\begin{figure}[t]
\centering
\includegraphics[width=0.78\textwidth]{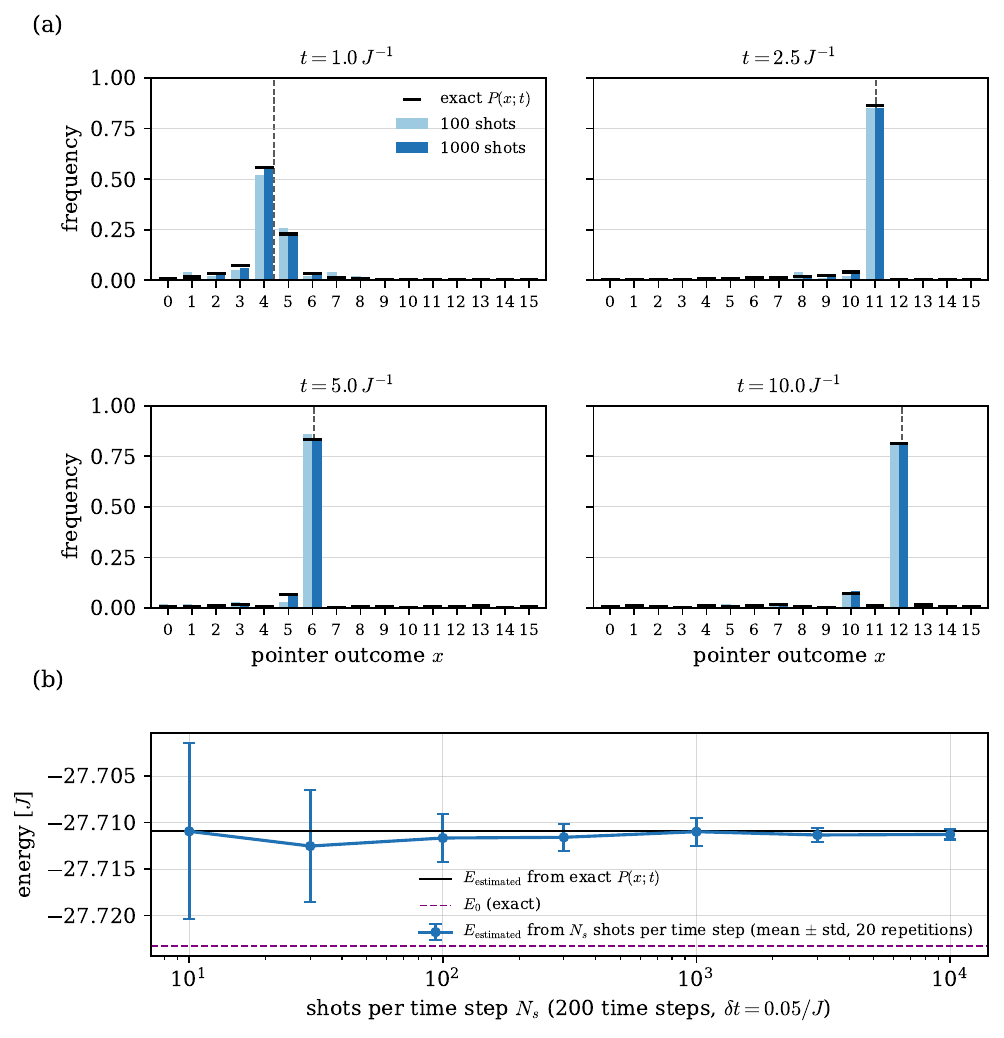}
\caption{Finite-shot pointer readout for the Heisenberg model on the $4\times3$ triangular lattice (data of Fig.~\ref{fig:heisenberg}: $N=12$, $r=4$, $\chidmrg=20$, $\delta t=0.05/J$, $\chitdvp^{\max}=128$). (a)~Frequencies of the pointer outcomes $x=0,\dots,15$ (computational basis after the inverse QFT) from $N_s=100$ (light) and $N_s=1000$ (dark) samples at $t=1$, $2.5$, $5$, and $10\,J^{-1}$; black marks: exact $P(x;t)$; dashed line: peak position $x^\star=-E_0t/2\pi \bmod 2^r$ for the exact ground-state energy. (b)~Energy estimate $E_{\mathrm{estimated}}$ obtained from $N_s$ samples at each of the 200 time steps, mean and standard deviation over 20 independent repetitions; solid line: estimate from the exact $P(x;t)$; dashed line: exact ground-state energy $E_0$. Conclusion: with $N_s\gtrsim100$ samples per time step, the shot noise of the estimate is several times smaller than its systematic deviation from $E_0$.}
\label{fig:sampling}
\end{figure}

\paragraph*{Results.} Figure~\ref{fig:sampling}(a) shows histograms of $N_s=100$ and $1000$ outcomes at four times, together with the exact $P(x;t)$ of Fig.~\ref{fig:heisenberg}. Already with 100 shots the dominant outcome is unambiguous at all times; the fluctuations mainly affect the small side peaks. Figure~\ref{fig:sampling}(b) shows the energy obtained by applying the estimator of Sec.~\ref{sec:tn_readout} (one-parameter fit of each row, median over rows) to histograms of $N_s$ outcomes at each of the 200 time steps, for 20 independent repetitions. The estimates are centered on the value obtained from the exact distribution, $-27.7110\,J$, with a standard deviation of $9.5\times10^{-3}\,J$ for $N_s=10$, $2.6\times10^{-3}\,J$ for $N_s=100$, $1.5\times10^{-3}\,J$ for $N_s=1000$, and $5.4\times10^{-4}\,J$ for $N_s=10^4$. For $N_s\gtrsim100$, i.e.\ $2\times10^4$ circuit executions in total, the statistical error is therefore several times smaller than the deviation of $1.2\times10^{-2}\,J$ from the exact ground-state energy, which is set by the initial state and the finite evolution time rather than by sampling. The least-squares fit weights all time points equally; a fit that accounts for the multinomial statistics of the counts, e.g.\ maximum likelihood, could reduce the number of shots further.

\section{Suzuki--Trotter product-formula error}
\label{app:trotter_error_analysis}
This appendix explains why a compiled digital circuit may yield energies that differ from those of the target Hamiltonian. This effect is distinct from the pointer-readout resolution of Appendix~\ref{app:vN_error_analysis} and from the TDVP truncation error of the tensor-network benchmarks.

\subsection{Suzuki--Trotter product formula}
Let the target Hamiltonian be a sum of terms, $\hat H=\sum_{i=1}^{K}\hat H_i$. The exact evolution over a short step $\delta t$ is $e^{-i\hat H\delta t}$; in the first-order Suzuki--Trotter approximation it is replaced by
\begin{equation}
e^{-i\hat H\delta t}=\prod_{i=1}^{K}e^{-i\hat H_i\delta t}+\mathcal O(\delta t^2),
\end{equation}
where the error arises from the non-commutativity of the terms $\hat H_i$. Repeating this over $n=t/\delta t$ steps approximates $e^{-i\hat Ht}$. Reducing $\delta t$ decreases the error but increases the circuit depth; higher-order product formulas reduce the error further at the cost of a more involved circuit.

\subsection{Consequence for extracted energies}
Because the compiled circuit implements a finite-step approximation, i.e.\ evolution under an effective Hamiltonian $\widetilde H_{\delta t}=\hat H+\mathcal O(\delta t)$, the energies extracted from the pointer need not coincide with the eigenvalues of $\hat H$. This bias decreases as $\delta t$ is reduced.

\subsection{Interpretation of the benchmark figures}
The data of Figs.~\ref{fig:heisenberg}--\ref{fig:pyridine} are obtained from TDVP/MPS simulations of the coupled system--pointer dynamics and therefore do not quantify the product-formula error at the same nominal $\delta t$. The chemical-accuracy behavior observed there is a benchmark of the protocol within the tensor-network simulation model, not a certification of the accuracy of a compiled Suzuki--Trotter circuit. In this work, the Suzuki--Trotter construction is used only as a logical implementation model for gate counting; a quantitative study of product-formula error for the benchmark Hamiltonians is left for future work.

\section{Remark on excited states}
\label{app:excited}
The same tensor-network machinery can be extended to excited states by working in subspaces orthogonal to previously obtained states, e.g.\ via projected-Hamiltonian or constrained-optimization formulations. A standard projected-Hamiltonian construction is
\begin{equation}
\hat H^{\mathrm p}=\big(\hat I-\ket\Omega\!\bra\Omega\big)\,\hat H\,\big(\hat I-\ket\Omega\!\bra\Omega\big),
\label{eq:H_projected}
\end{equation}
where $\ket\Omega$ is the ground state. The terms of Eq.~\eqref{eq:H_projected} can be represented as MPOs, so that the ground state of $\hat H^{\mathrm p}$, i.e.\ the first excited state of $\hat H$, can be obtained with standard DMRG-type optimization~\cite{baiardi_density_2020, mcculloch_density-matrix_2007, Keller2015MPO}. The quality of higher excited states depends on the accuracy of the lower states used in the orthogonality constraints, so errors can accumulate in successive constrained optimizations. Other strategies, such as quantum subspace expansion, may provide useful alternatives. Since the present work focuses on ground-state benchmarks, we include this discussion only as an outlook and make no benchmark claims for excited states.

\bibliography{references_revised}
\bibliographystyle{apsrev4-2}

\end{document}